\documentclass[twocolumn]{aastex631}

\usepackage{amsmath}
\usepackage{listings}

\newcommand\sysname{\texttt{RADAR}}

\begin{document}

\title{RADAR---Radio Afterglow Detection and AI-driven Response: A Federated Framework for Gravitational Wave Event Follow-Up}

\author[0000-0002-9043-7059]{Parth Patel}
\affiliation{Data Science and Learning Division,
Argonne National Laboratory, Lemont, IL 60439, USA}

\author[0000-0001-8104-3536]{Alessandra Corsi}
\affiliation{William H. Miller III Department of Physics and Astronomy,
Johns Hopkins University, Baltimore, Maryland 21218, USA}

\author[0000-0002-9682-3604]{E.~A. Huerta}
\affiliation{Data Science and Learning Division,
Argonne National Laboratory, Lemont, IL 60439, USA}
\affiliation{Department of Computer Science, The University of Chicago, Chicago, Illinois 60637, USA}
\affiliation{Department of
Physics, University of Illinois Urbana-Champaign,
Urbana, IL 61801, USA}
\affiliation{Department of
Astronomy, University of Illinois Urbana-Champaign,
Urbana, IL 61801, USA}

\author[0000-0003-1773-5372]{Kara Merfeld}
\affiliation{William H. Miller III Department of Physics and Astronomy,
Johns Hopkins University, Baltimore, Maryland 21218, USA}

\author[0000-0001-8544-6498]{Victoria Tiki}
\affiliation{Data Science and Learning Division,
Argonne National Laboratory, Lemont, IL 60439, USA}
\affiliation{Department of
Physics, University of Illinois Urbana-Champaign,
Urbana, IL 61801, USA}

\author[0000-0002-0160-2519]{Zilinghan Li}
\affiliation{Data Science and Learning Division,
Argonne National Laboratory, Lemont, IL 60439, USA}

\author[0000-0002-8428-5159]{Tekin Bicer}
\affiliation{Data Science and Learning Division,
Argonne National Laboratory, Lemont, IL 60439, USA}

\author[0000-0002-7370-4805]{Kyle Chard}
\affiliation{Data Science and Learning Division,
Argonne National Laboratory, Lemont, IL 60439, USA}
\affiliation{Department of Computer Science, The University of Chicago, Chicago, Illinois 60637, USA}

\author[0000-0002-6781-7432]{Ryan Chard}
\affiliation{Data Science and Learning Division,
Argonne National Laboratory, Lemont, IL 60439, USA}

\author[0000-0002-9682-3604]{Ian T.~ Foster}
\affiliation{Data Science and Learning Division,
Argonne National Laboratory, Lemont, IL 60439, USA}
\affiliation{Department of Computer Science, The University of Chicago, Chicago, Illinois 60637, USA}

\author[0000-0003-3306-0483]{Maxime Gonthier}
\affiliation{Data Science and Learning Division,
Argonne National Laboratory, Lemont, IL 60439, USA}
\affiliation{Department of Computer Science, The University of Chicago, Chicago, Illinois 60637, USA}

\author[0000-0002-4830-4535]{Valerie Hayot-Sasson}
\affiliation{Data Science and Learning Division,
Argonne National Laboratory, Lemont, IL 60439, USA}
\affiliation{Department of Computer Science, The University of Chicago, Chicago, Illinois 60637, USA}

\author[0000-0003-4177-0493]{Hai Duc Nguyen}
\affiliation{Data Science and Learning Division,
Argonne National Laboratory, Lemont, IL 60439, USA}

\author[0009-0006-8992-5895]{Haochen Pan}
\affiliation{Department of Computer Science, The University of Chicago, Chicago, Illinois 60637, USA}

\begin{abstract}
The landmark detection of both gravitational waves (GWs) and electromagnetic (EM) radiation from the binary neutron star merger GW170817 has spurred efforts to streamline the follow-up of GW alerts in current and future observing runs of ground-based GW detectors. Within this context, the radio band of the EM spectrum presents unique challenges. Sensitive radio facilities capable of detecting the faint radio afterglow seen in GW170817, and with sufficient angular resolution, have small fields of view compared to typical GW localization areas. Additionally, theoretical models predict that the radio emission from binary neutron star mergers can evolve over weeks to years, necessitating long-term monitoring to probe the physics of the various post-merger ejecta components. These constraints, combined with limited radio observing resources, make the development of more coordinated follow-up strategies essential---especially as the next generation of GW detectors promise a dramatic increase in detection rates. Here, we present \sysname{}, a framework designed to address these challenges by promoting community-driven information sharing, federated data analysis, and system resilience, while integrating AI methods for both GW signal identification and radio data aggregation. We show that it is possible to preserve data rights while sharing models that can help design and/or update follow-up strategies. We demonstrate our approach through a case study of GW170817, and discuss future directions for refinement and broader application.
\end{abstract}


\section{Introduction} \label{sec:intro}
The field of multi-messenger astrophysics (MMA) has been spearheaded by the discovery of sources of high-energy neutrinos \citep{2014PhRvL.113j1101A,2018Sci...361.1378I,2022Sci...378..538I} and gravitational waves \citep[GWs;][]{2016PhRvL.116f1102A}, and their electromagnetic (EM) counterparts \citep[e.g.,][and references therein]{2017PhRvL.119p1101A,2017ApJ...848L..12A,2017Sci...358.1556C,2017Natur.551...80K,2017Sci...358.1559K,2017Natur.551...71T,margutti2018binary,2017Natur.551...67P,2018Sci...361..147I,2022Sci...378..538I}. Radio observations of the binary neutron star merger GW170817, and their extensive modeling, have revealed the first off-axis structured relativistic jet from a gamma-ray burst (GRB) observed about 2\,s after the merger \citep[e.g.,][]{2017ApJ...848L..13A,2017ApJ...848L..21A,2017ApJ...848L..14G,2017Sci...358.1579H,2017ApJ...848L..14G,dobie2018turnover, 2018MNRAS.473..576G,2018Natur.554..207M,2018Natur.561..355M,2018PhRvL.120x1103L,2019Sci...363..968G}.
Cosmic neutrino events such as IceCube-170922A, from the known blazar TXS 0506+056, have established a link between high-energy neutrinos and supermassive black holes in active galactic nuclei \citep[AGN;][]{2018Sci...361.1378I,2018Sci...361..147I,2021A&A...650A..83H,2021PhRvD.103l3018Z,2023arXiv230804311E,2024ApJ...973...97A,2024A&A...690A.111K}.  Radio-quiet AGN have also emerged as candidate EM counterparts to high-energy neutrino events \citep[e.g.,][]{2004PhRvD..70l3001A,2022ApJ...941L..17M,2023ApJ...956....8F,2024PhRvD.110l3014K,2025MNRAS.tmp..974M}.

In this work, we focus on multi-messenger studies of stellar-mass compact object mergers. These mergers are important targets for ground-based GW detectors such as LIGO-Virgo-KAGRA \citep{2015CQGra..32g4001L,2015CQGra..32b4001A,2021PTEP.2021eA101A}. As demonstrated by the case of GW170817 \citep{2017Sci...358.1579H,2021ApJ...922..154M}, radio observations of mergers can complement what is learned from GWs by probing the presence of fast, non-thermally emitting merger ejecta that cannot be observed at optical wavelengths. While $\gamma$-rays and X-rays can also be used to study fast ejecta \citep{Fong2015,margutti2018binary,hajela2019two,2025MNRAS.539.2654K}, the radio band is unique in its ability to probe emissions
largely independently of geometric effects, and from ejecta components with speeds varying from mildly- to ultra-relativistic \citep[][]{2018ApJ...867...95H,2018ApJ...868L..11M,2018Natur.554..207M,2021ApJ...914L..20B,2022ApJ...938...12B,2024FrASS..1101792C}. The radio band also enables very-long baseline interferometry (VLBI), via which proper motion of source structures can be measured \citep{2018Natur.554..207M,2019Sci...363..968G}, with impact on standard siren cosmology and the measurement of the Hubble constant $H_0$ \citep{2019NatAs...3..940H}. Radio polarization measurements can also constrain the magnetic field structure associated with relativistic jets \citep[e.g.,][and references therein]{2018ApJ...861L..10C,2018MNRAS.478.4128G}.  

Over the next five years,  the relatively small field of view (FOV) of sensitive and high-resolution
radio interferometers operating in the GHz domain is likely to remain the major challenge for radio follow-up observation of GW events. Indeed, over this timescale, the median GW localizations of compact binary mergers of neutron stars are likely to remain $\gtrsim 100$\,deg$^2$ \citep{2018LRR....21....3A}. Hence,
the hunt for radio counterparts will have to rely on the identification and localization at other wavelengths by larger FOV instruments \citep{2022ApJS..260...18A}. In this context, as demonstrated by the massive observational effort that followed the GW170817 discovery \citep{2017ApJ...848L..12A}, community sharing of
critical information (sky position, spectral classification, temporal behavior) on
candidate GW counterparts \citep[e.g.,][]{2020ApJ...894..127W} is key to enabling radio follow-up observations and their optimization.
Moreover, radio emission in compact binary mergers can arise from a variety of ejecta components (jet wings, jet core, dynamical ejecta tail) with different geometries (on- or off-axis) and with characteristic emission peaks ranging from early times (potentially even seconds to minutes before the merger), to very late times (years after the merger), hence requiring both fast follow-up capabilities and extended resources (telescope time) for long-term observations \citep{2010Ap&SS.330...13P,2011Natur.478...82N,2015ApJ...806..224M,2016ApJ...831..190H,2021MNRAS.501.3184S,2021MNRAS.505.2647D,2024FrASS..1186748C,2024arXiv241010579K,2025arXiv250319884S}. So, again, community sharing of early-time radio follow-up results can benefit later-time follow-up strategies and help optimize the use of telescope time over the long timescales required to probe the physics of the mergers. 

Here, we present \texttt{RADAR}---\texttt{R}adio \texttt{A}fterglow \texttt{D}etection and \texttt{A}I-driven \texttt{R}esponse, a framework for radio follow-up of GW events that enables collaborative, model-informed observing strategies while accommodating heterogeneous data-sharing policies. \texttt{RADAR} is not intended to replace current or future low-latency GW pipelines developed and deployed within the LIGO-Virgo-KAGRA collaboration \citep[e.g.,][and references therein]{2024PNAS..12116474C}. Rather, it is a framework complimentary to LIGO-Virgo-KAGRA and other community efforts \citep[e.g.,][]{2012SPIE.8448E..0QS,2012A&A...539A.124L,2018LRR....21....3A,2019ApJ...875..161A,2019PhRvD.100j3025C,2019PhRvD.100f3015G,2020ApJ...905L..25S,2021ApJ...910L..21M,2021PhRvD.104b3014M,2021PhRvD.103j3006W,2022NatPh..18..112G,2023ApJ...958L..43H,2024CQGra..41h5012B,2024MLS&T...5d5030C,2024PhRvD.109d2008E,2024ApJ...963...98R,2024EPJWC.29504022V,2025LRR....28....2C,2025Natur.639...49D,2025PhRvD.111d2010M}. Instead, \texttt{RADAR} explores how site-local AI inference and federated coordination mechanisms can support cross-observatory workflows under real-world constraints, particularly in the radio domain. The motivation for \texttt{RADAR} stems from the recognition that while centralized alert systems and optical brokers have made significant progress in coordinating follow-up \citep[e.g.,][]{2019PASP..131a8001P,2000AIPC..526..731B,2022GCN.32419....1B,2019A&A...631A.147N,2021AJ....161..242F,2021AJ....161..107M,2023GCN.33638....1B,2023ApJS..267...31C,2024ApJ...964...35H,2023PASP..135f4501C,2024FrASS..1101785H}, the radio community faces distinct challenges that we have already underlined: from narrow fields of view, to extended follow-up needs where rapid contextualization in the form of early radio data modeling may help guide whether long-term monitoring is warranted in  resource-constrained settings. \texttt{RADAR} addresses this by enabling model-informed, asynchronous follow-up planning across distributed facilities, without requiring centralized data transfer or early public release. In fact,  proprietary periods vary significantly across facilities, ranging from immediate public release (e.g., some space-based missions) to restricted-access periods in the case of, e.g., LIGO and U.S.\  national radio observatories. Data sharing practices also vary among user groups, with some teams releasing measurements promptly without accompanying data, and others delaying dissemination until peer-reviewed publication. Leveraging a privacy-enhancing 
federated framework, we perform federated learning on proprietary, site-local high-level observables (e.g., flux densities) without transferring raw data. Although \sysname{} does not implement 
privacy-preserving federated learning~\citep{ZL11044858}, 
it is guided by its underlying principles. Rather than training shared AI models~\citep{li2024secure}, we coordinate light curve fitting through global parameter exchange, facilitating collaborative afterglow characterization while preserving data locality, safeguarding privacy, and respecting institutional data ownership. In this sense, \sysname{} adopts a privacy-enhancing framework, 
and enables coordinated, data-informed follow-up campaigns while preserving each group’s autonomy to publish complete analyses, which typically contain data reduction procedures, imaging, localization, spectral energy distribution analysis, and more comprehensive modeling. 

As a case study, we apply \texttt{RADAR} to GW170817. { At the time of writing, GW170817 remains the only binary neutron star merger detected in GWs with a secure EM counterpart identification. GW170817 enjoys a rich MMA dataset which allows controlled, reproducible testing of \texttt{RADAR} across diverse data contributors. Hence, this event serves as a proof-of-concept validation of RADAR’s architecture, including its AI modules, federated orchestration, and MMA integration.} Although in the case of GW170817 some key observables (e.g., counterpart position and flux density at discovery) were made promptly available to the community, we show that a federated, privacy-enhancing framework that also incorporates proprietary observations could have enabled a more automated and coordinated effort.  Overall, the approach proposed here facilitates broad participation in radio follow-up efforts, even under diverse data-sharing practices, and is readily extendable to other messengers and EM bands. Indeed, \texttt{RADAR} framework’s design allows its modules, whether for GW signal detection, parameter estimation, or radio light curve modeling, to be easily substituted or extended, enabling integration with state-of-the-art developments in each subdomain. Hence, \texttt{RADAR}'s major contribution lies not in proposing new detection models per se, but in integrating key components such as detection, inference, and follow up within a privacy-aware, federated workflow that respects data ownership and adapts to heterogeneous infrastructure. By demonstrating that these components can work together in practice across supercomputing centers, private observatories, and public alert systems, \texttt{RADAR} fills a gap in the current MMA ecosystem: a lightweight, extensible architecture for collaborative science under evolving data-sharing norms.

\begin{figure*}
\begin{center}
\includegraphics[width=0.95\textwidth,trim=1mm 1mm 1mm 1mm,clip]{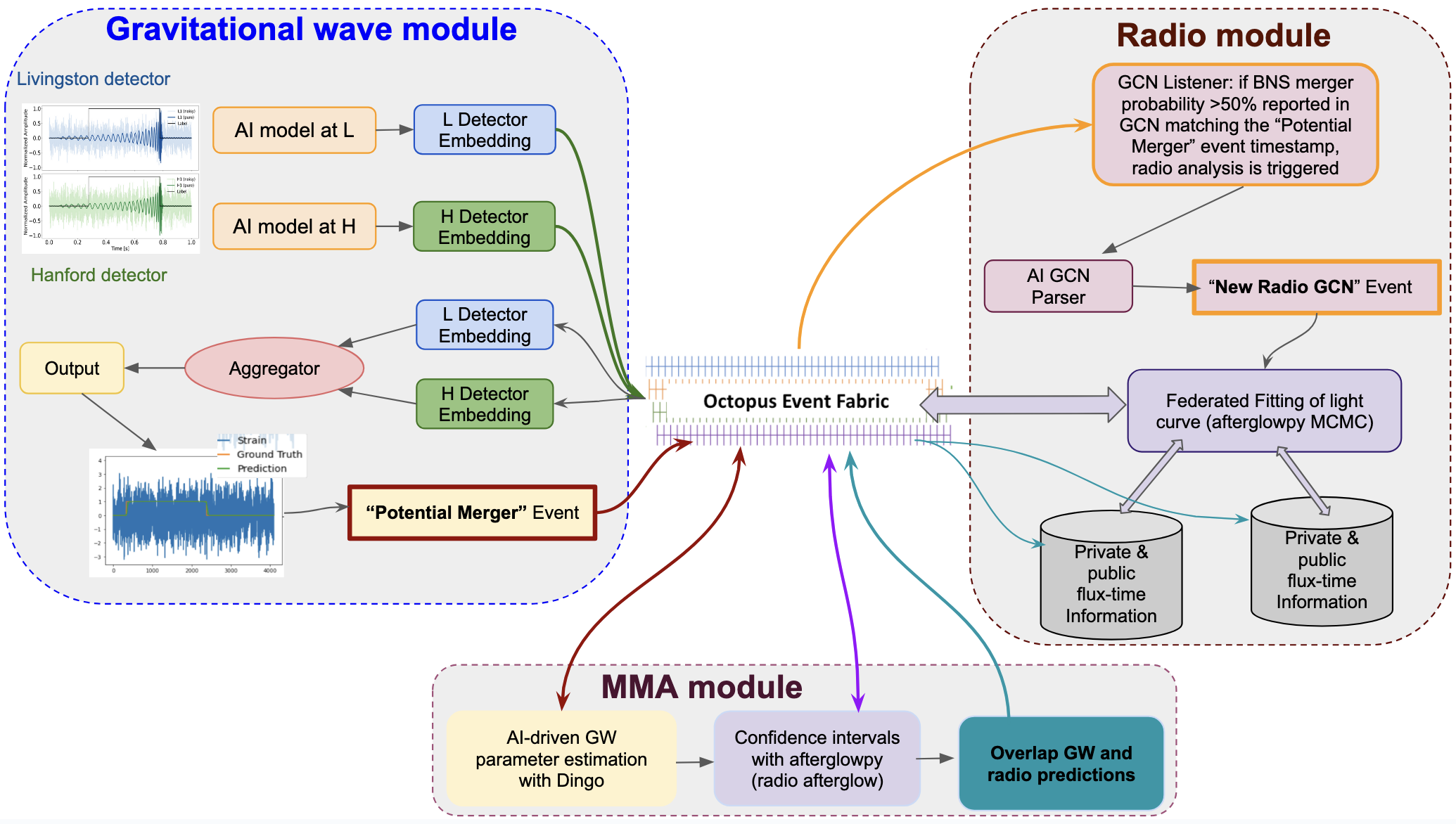}\caption{Schematic overview of \texttt{RADAR}, our federated, AI-driven computational framework for multi-messenger astrophysics (MMA). \texttt{RADAR} consists of four interconnected modules that collectively enable detection, federated inference, and joint scientific interpretation of GW and EM data. For this demonstration, rather than interfacing directly with the LIGO Hanford and Livingston detectors, we emulate detector endpoints by hosting their respective data streams on two distinct HPC platforms---the DeltaAI supercomputer at the National Center for Supercomputing Applications (NCSA) and the Polaris supercomputer at the Argonne Leadership Computing Facility (ALCF). \textit{(1)} The \textit{Gravitational Wave module} (left) deploys AI models at each site, DeltaAI and Polaris, for prompt signal detection and latent embedding generation. Embeddings are then transmitted to a central server hosted on the Delta supercomputer at NCSA for post-processing and candidate event identification. In the graph at the bottom left of the box, we show GW data containing a signal (``Ground Truth'') correctly identified by the AI model. The corresponding ``Prediction'' is indicated by the line that spikes to 1 when the signal is detected and returns to 0 otherwise. See also \autoref{fig:fl_signal} and \autoref{sec:appendix_model} for more details. The \textit{Radio module} (right) performs federated light curve modeling using \texttt{afterglowpy}, aggregating radio follow-up data from both public GCN circulars and private observatories. A dedicated \textit{GCN listener} ingests new notices in real time, while an \textit{AI GCN Parser}, based on the \texttt{GPT-4.1} LLM, extracts structured metadata and flags incomplete entries to ensure robust downstream analysis. \textit{(3)} The \textit{MMA module} (bottom) performs posterior consistency checks by comparing overlapping parameters (e.g., luminosity distance $D_L$ and viewing angle $\theta_{\rm obs}$) inferred from GW data, through posterior sampling via the \texttt{Dingo-BNS} inference engine, and EM channels, through posterior sampling via \texttt{afterglowpy}, leveraging federated posteriors published by upstream modules. \textit{(4)} The \textit{Octopus Event Fabric} (center) acts as a messaging and coordination layer, enabling asynchronous communication of posterior samples and intermediate results from producers to consumers for progressive inference. These mechanisms allow the overall architecture to support data locality, minimize communication overhead, track provenance, and achieve privacy-enhancing collaborative analysis across heterogeneous facilities.}\label{fig:workflow}
\end{center}
\end{figure*}

As we progress toward an era of next generation radio \citep[e.g.,][]{2019BAAS...51g.255H,2018ASPC..517....3M,2025arXiv250106333C} and GW facilities \citep{2021arXiv210909882E,2024CQGra..41x5001G,2023arXiv230613745E,2025arXiv250312263A}, radio follow-up efforts are likely to benefit from both improved GW sky localizations and faster radio survey speeds \citep{2025arXiv250622835M}. However, the anticipated increase in GW detection rates—by approximately a factor of $10^3$ compared to current levels \citep{2021arXiv210909882E,2024CQGra..41x5001G,2025arXiv250312263A,2019NatRP...1..600H}—means that methods supporting community-wide information sharing for planning observational strategies and reducing redundant efforts (such as the one presented here) will continue to play a vital role \citep{2024arXiv240102063T}.

The remainder of this paper is organized as follows. In \autoref{sec:methods}, we introduce \texttt{RADAR}. \autoref{sec:gw-module} and \autoref{sec:radio-module} detail the implementation of \texttt{RADAR}'s components responsible for GW and radio data processing, respectively. In \autoref{sec:mma-module}, we describe the methodology used to integrate GW and radio data in a joint multi-messenger analysis. \autoref{sec:octopus} outlines the orchestration layer that links the major components of the system. In \autoref{sec:performance}, we evaluate the end-to-end performance of \texttt{RADAR} using GW170817 as a case study. \autoref{sec:resilience} presents an initial assessment of the robustness and fault tolerance of the key computational modules. We conclude with a summary and discussion of future directions in \autoref{sec:conclusion}.

\section{Computational framework} \label{sec:methods}
\sysname{} has four main components (\autoref{fig:workflow}): 
\begin{enumerate}
\item \textit{The GW module}, represented by the left box in \autoref{fig:workflow}, is composed of an updated version of the AI models for signal detection previously presented in \cite{Tian:2023vdc}. Specifically, AI models are sent to the emulated data sources, where LIGO Hanford and Livingston data streams are hosted on the DeltaAI supercomputer at NCSA and the Polaris supercomputer at ALCF, respectively, and upon 
processing GW data, embeddings (i.e., vector representations of information) are sent to a central server on the Delta 
supercomputer at NCSA where post-processing is done in search for GW signals. We note that this particular choice of GW detection algorithm is intended solely for demonstration purposes; alternative GW detection pipelines could be integrated within the \texttt{RADAR} framework, as discussed in Section~\ref{sec:intro}. If a GW signal is identified, its 
GPS time is recorded, compared with that of any public LIGO-Virgo-KAGRA alerts, and the second component of the analysis is 
triggered (see below). 
\item \textit{The radio module}, represented by the right box in \autoref{fig:workflow}, aims to aggregate public (e.g., data contained in public notices such as GCNs) and private data collected by a variety of users to estimate key physical parameters of the EM counterparts potentially associated with the GW signals identified via the GW module. As a test case, we use \texttt{afterglowpy} \citep{Ryan_2020}, a Python module to calculate afterglow light curves (and spectra) of relativistic jets associated with compact binary mergers. 
\item \textit{The MMA module}, represented by the bottom box in \autoref{fig:workflow}, aims to derive constraints on the physics of compact binary mergers combining \texttt{Dingo-BNS} \citep{2021PhRvL.127x1103D}, a Python program 
for analyzing GW data using neural posterior estimation, and \texttt{afterglowpy} (see above). 
Here, we focus on combining GW and radio data specifically for the purpose of improving our estimates of compact binary mergers luminosity distances and viewing angles (i.e., the angles between the observers and the polar axis of the binaries). These estimates impact fundamental physics questions such as the  Hubble tension \citep[e.g., ][and references therein]{2017Natur.551...85A,2019NatAs...3..940H,2025ApJ...979L...9S}. However, the joint MMA analysis can be generalized to any set of parameters that can be constrained via both GW and EM data.
\item \textit{The \texttt{Octopus} event fabric} \citep{pan2024octopus} is used to publish  
information regarding the status of key computations throughout 
\sysname{}, as well as information regarding key physical parameters of the MMA sources (\autoref{fig:workflow}, center).
\end{enumerate}

In the sections that follow, we describe each component in detail. 

\section{The GW module}
\label{sec:gw-module}

\noindent Traditional GW inference methods often 
involve moving GW data 
to a central location for computationally expensive 
post-processing~\citep{2019PASP..131b4503B,CANNON2021100680}. 
Both the transfer and processing costs can introduce delays in MMA follow-up observations. 
Here, we present an improved version of the AI models for GW detection introduced in~\cite{Tian:2023vdc} 
(see \autoref{sec:appendix_model} for a more detailed description). 
In this approach, AI models are communicated to each of the two LIGO sites to process 
data in-situ, and then the output of these AI models is sent to 
a central server where it is aggregated and post-processed to 
identify GW signals. 
The distributed architecture preserves data locality 
while maintaining the ability to detect and 
characterize potential merger events across both detector sites through 
AI model components. These AI model components are:  
a hierarchical dilated convolution neural network (HDCN) to 
capture temporal features in Livingston and Hanford data, and 
an aggregator module that produces a unified prediction using the HDCN output from each detector. 

\begin{figure}
    \centering    \includegraphics[width=\columnwidth]{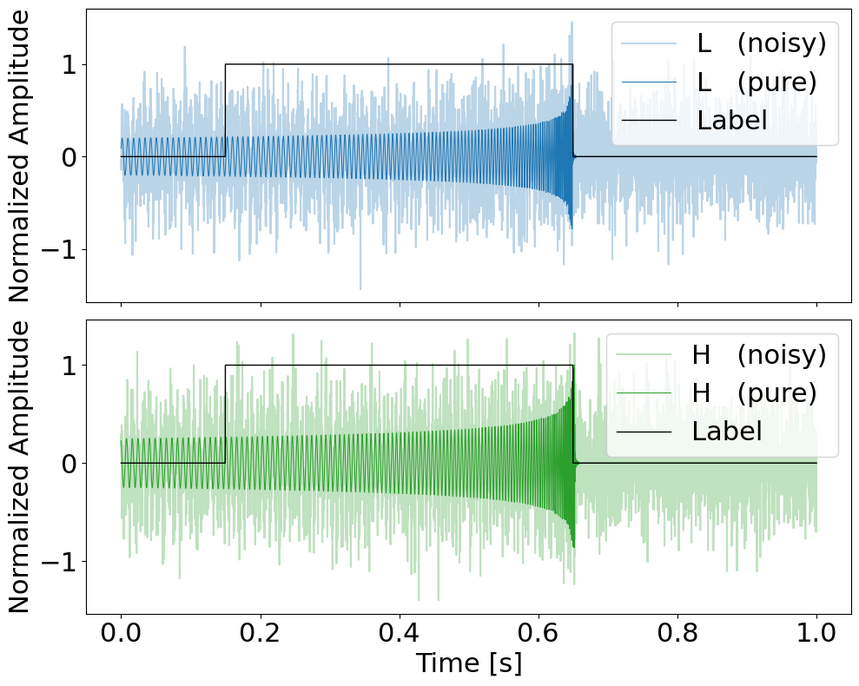} 
    \caption{Whitened noisy gravitational wave signal for Livingston (L) and Hanford (H) signals. For training the AI model used for GW detection, each waveform is labeled ``1'' in the 0.5\,s leading up to the merger event and ``0'' elsewhere. See text for discussion. }
    \label{fig:fl_signal}
\end{figure}

We 
use the HDCN to generate embeddings from the local strain data at 
a LIGO detector site. 
By treating each detector's data as independent and identically 
distributed, we ensure that the embeddings retain site-specific 
information, which is crucial for subsequent aggregation. 
The HDCN employs dilated causal convolutions 
that allow the network to capture efficiently the long-range 
temporal dependencies characteristic of GW signals. The locally 
generated embeddings are transmitted to a central server where 
they are aggregated using the aggregator module (``Aggregator'' in \autoref{fig:workflow}, left). The server processes the aggregated 
embeddings using time-independent operations that enable 
the model to generate a score for each time step, indicating 
how similar the input is to a segment containing a GW signal. 

Upon completion of the above inference phase, a 
post-processing pipeline is initiated to identify potential 
GW events. This pipeline uses \texttt{scipy.signal.find\_peaks} 
to identify peaks that exceed a predefined threshold and span 
a minimum width, filtering out short or low-confidence 
detections. Only peaks with sufficiently high average scores 
across the region are retained and return a trigger. For each 
validated detection, the pipeline generates  both GPS 
and UTC timestamps. These timestamps are crucial for enabling 
prompt follow-up observations across the electromagnetic 
spectrum, facilitating MMA studies (\autoref{sec:octopus}). 

Our AI models for GW detection are trained 
using modeled waveforms sampled at 4096\,Hz, 
and GW data, sampled at 4096\,Hz, available at the 
Gravitational Wave Open Science Center \citep{2021SoftX..1300658A,2023ApJS..267...29A}. Power Spectral Densities 
(PSDs) are estimated using real LIGO data, and then used 
to whiten LIGO noise and modeled waveforms. LIGO noise 
used for this work does not contain known detections. 
Thereafter, whitened noise and whitened waveforms 
are linearly combined to train AI models with waveforms 
that have a broad range of signal-to-noise 
ratios. We used independent, non-overlapping sets of linearly 
combined whitened noise and whitened waveforms for training, 
validation and testing.  
Each waveform is labeled ``1'' in the 0.5\,s leading 
up to the merger event and ``0'' elsewhere 
(see \autoref{fig:fl_signal}). 
For inference, the model processes 1\,s data segments 
(4096 samples) and applies a peak detection algorithm, 
identifying GW signals based on specific height (threshold) 
and width requirements to separate signals from noise.

As we describe in more detail in \autoref{sec:octopus}, 
the identification of a GW event via the GW module represents 
the first step in \sysname{}. The event's timestamp is compared with 
the time of any public LIGO-Virgo-KAGRA alerts. When a time 
coincidence is identified with a LIGO-Virgo-KAGRA alert 
consistent with a binary neutron star merger, a ``trigger'' 
alert is generated and other critical steps (radio and MMA 
analysis; Sections~\ref{sec:radio-module} and~\ref{sec:mma-module}) 
in the framework are activated.

\begin{table*}
\centering
\caption{The parameters used to model the radio emission from relativistic jet afterglows in \texttt{afterglowpy} (columns 1 and 2). In column 3 we report the ranges adopted as priors for the \texttt{afterglowpy} fits that we use in \sysname{} for the GW170817 case study. Since we consider only radio data here, we fix the jet structure and the microphysics parameters to the multi-wavelength best fit values reported in \citet{Troja_2020}. We also fix the redshift $z$ to that of the host galaxy of the GW170817 optical counterpart (kilonova) host galaxy \citep{2017ApJ...848L..31H}. We then fit for $E_0$, $n_0$, $D_L$ and $\theta_{\rm obs}$. We adopt very broad priors for $E_0$ and $n_0$   \citep[including the most extreme values reported in][for the observed population of short GRBs]{Fong2015}. For $D_L$ and $\theta_{\rm obs}$ we adopt as priors the 68\% posterior obtained from the GW data using \texttt{Dingo-BNS}. \label{tab:AfterglowpyParams} }
\begin{tabular}{ccc}
\hline\hline
Symbol & Description & Parameter Space\\
\hline
jetType & $\theta$-dependent jet structure & Gaussian\\
\hline
$\theta_w$ & Truncation angle & Fixed at 0.6\,rad \\
\hline
$E_0$ & Isotropic energy equivalent & $10^{48}$ -- $10^{54}$ erg \\
\hline
$\theta_c$ & Half-opening angle & Fixed at 0.088\,rad \\
\hline
$\theta_{obs}$ & Observing angle & Provided by \texttt{Dingo-BNS} \\
\hline
$n_{0}$ & ISM density & $10^{-5}$--$10^{2}$ $\text{cm}^{-3}$ \\
\hline
$p$ & Electron energy distribution index & Fixed at 2.139\\
\hline
$\epsilon_e$ & Fraction of energy given to electrons & Fixed at 0.01 \\
\hline
$\epsilon_B$ & Fraction of energy given to magnetic field & Fixed at 0.0002\\
\hline
$z$ & Redshift & Fixed at 0.0098\\
\hline
$D_L$ & Luminosity distance & Provided by \texttt{Dingo-BNS}\\
\hline
\hline
\end{tabular}
\end{table*}

\section{The Radio module}
\label{sec:radio-module}
We assume in this work that radio follow-up observations of GW events or event candidates are to be performed following the identification of an EM counterpart or EM counterpart candidate localized to arcsec precision (within the generally much larger GW localization areas) via observations at wavelengths other than radio. This assumption reflects the strategy that is most commonly adopted for EM follow up by the community, in light of the limitations in the localization capabilities of the current generation of ground-based GW detectors. In \autoref{sec:conclusion}, we discuss how \sysname{} could be modified with the advent of next-generation GW detectors potentially enabling more accurate GW localizations.

\subsection{Radio data aggregation}
\label{sec:radiodataaggregation}
The first step in the radio module is to collect radio data from both public sources, such as the GCN alerts, and non-public sources, such as observers storing proprietary observations at various sites. 

GCN alerts are obtained by a GCN ``listener'' that listens continuously to public LIGO-Virgo-KAGRA notices and GCN circulars. Once a Trigger is received (as we describe in \autoref{sec:octopus}, a Trigger is represented by a time coincidence between an event from the GW module and a likely binary neutron star merger in a LIGO-Virgo-KAGRA notice), the system starts evaluating incoming notices and circulars for references to known GW events by searching for event identifiers (e.g., GW170817) within the title, event ID field, or body text. Relevant notices and circulars are saved in appropriate event folders. 

Circulars pertaining to a specific GW event are further processed to search for information on radio follow up observations. Once a radio circular is identified, a specialized AI interpreter processes the unstructured circular text into machine-readable data suitable for scientific analysis, facilitating rapid follow-up and reducing manual processing overhead. Our AI-powered GCN parser employs a large language model (LLM; specifically, \texttt{ChatGPT}'s \texttt{GPT 4.1} model) to extract from the radio circulars the information needed to construct a light curve: name of the observed source, Right Ascension and Declination, time observed, frequency of the observation, measured flux density, and flux density error. The extracted information is collated in an output table that is fed as input data to \texttt{afterglowpy}. 

The system also listens for data from multiple geographically distributed sites (represented by the gray cylinders in the radio module box of \autoref{fig:workflow}) that can then be used for privacy-enhancing light curve fitting, as described in the next Section. 

\subsection{Radio light curve modeling}
We use \texttt{afterglowpy} \citep{Ryan_2020}, a software package for modeling multi-wavelength light curves of synchrotron-emitting relativistic jets, such as those powering GRBs associated with the coalescences of compact binary systems involving at least one neutron star \citep[see e.g.][ for a review]{2014ARA&A..52...43B}. 

As the case of GW170817 highlighted, the jet structure (i.e., the jet energy distribution as a function of polar angle from the jet axis) is not well known. In \texttt{afterglowpy}, the jet is hence assumed to be described by a top-hat (constant energy density within the jet opening angle), Gaussian, or power-law energy profile (see \autoref{tab:AfterglowpyParams}). Hereafter, we assume a Gaussian jet of the form: 
\begin{equation}   
E_{\rm Gaussian}(\theta) = E_0 \exp{\left( -\frac{\theta ^2}{2\theta_c^2}\right)} .
\label{Eq:Power-Law}
\end{equation}
The parameters that define the model are the viewing angle $\theta_{\rm obs}$; $E_0$, the on-axis isotropic equivalent energy in ergs;
$\theta_{\rm c}$, the half-width of the jet core; $\theta_{\rm w}$, the ``wing'' truncation angle of the jet;
$n_0$, the number density of the ISM in cm$^{-3}$; $p$, the electron distribution power-law index; $\epsilon_e$, the fraction of thermal energy in electrons; $\epsilon_B$, the thermal energy fraction in magnetic fields; $d_L$, the luminosity distance; and $z$, the redshift.
    
The \texttt{afterglowpy} code can be used to perform a Markov chain Monte Carlo (MCMC) analysis in the parameter space defined in \autoref{tab:AfterglowpyParams} via the \texttt{emcee} routine, an MIT licensed pure-Python implementation of \citet{2010CAMCS...5...65G}'s Affine Invariant MCMC Ensemble sampler. To this end, we developed a script similar to that presented in \citet{Eyles-Ferris_2024}, but with modifications to account for, e.g., radio bands and exact observing frequencies.
Example results of the light curve fitting and parameter estimation procedures for GW170817 are presented in Figures~\ref{fig:GW170817_lightcurve} and~\ref{fig:posteriors_side_by_side}, and discussed in detail in \autoref{sec:performance}.

To enable privacy-enhancing radio light curve modeling, our system ensures that sensitive astronomical data remains at each participating site while still enabling collaborative light curve fitting across multiple geographically distributed locations. Privacy enhancement is achieved through a federated learning approach where raw observational data never leaves its originating site, yet all sites can collectively contribute to fitting light curve parameters. This is accomplished through a secure communication infrastructure built on Octopus and ProxyStore. Octopus provides robust authentication and authorization mechanisms by bridging Globus Auth identities with AWS IAM identities, where users authenticate to Octopus using Globus Auth credentials that are then translated into appropriate IAM credentials. This integration leverages AWS's native security features and enables fast, configurable access control on a per-topic basis for Octopus topics, ensuring that only authorized participants can access specific data streams or computational results. ProxyStore complements this by enabling secure data movement through a proxy-based architecture where lightweight proxy objects can be moved in place of confidential astronomical data, ensuring that the actual data are only resolved and accessed where permitted by the security policies. Together, these components ensure that individual sites cannot access raw data from other participating locations while still enabling collective computation of light curve parameters, thus preserving data locality and privacy while facilitating the scientific collaboration necessary for accurate astronomical modeling.

\section{The MMA module} \label{sec:mma-module}
The MMA module (bottom box in Figure \ref{fig:workflow}) is designed to derive combined constraints on parameters of the source (binary neutron star merger) using the GW and EM (radio) data. In this work, we emphasize the ability of radio observations to reduce the degeneracy between the luminosity distance $D_L$ and the observing  angle $\theta_{\rm obs}$ (i.e., the polar angle between the observer's line of sight and the perpendicular to the plane containing the merging neutron star system) that affects GW observations. As already demonstrated by \citep{2019NatAs...3..940H} (see also references therein), reducing this degeneracy has important consequences, such as, for our ability to use MMA observations to derive standard siren constraints on the Hubble constant $H_0$. 

The MMA module runs a GW parameter estimation pipeline named \texttt{Dingo-BNS} \citep{Dax:2024mcn}, a machine learning tool  designed for fast and accurate inference of GW signals from binary neutron star mergers. 
\begin{figure}
    \centering
\includegraphics[width=\columnwidth]{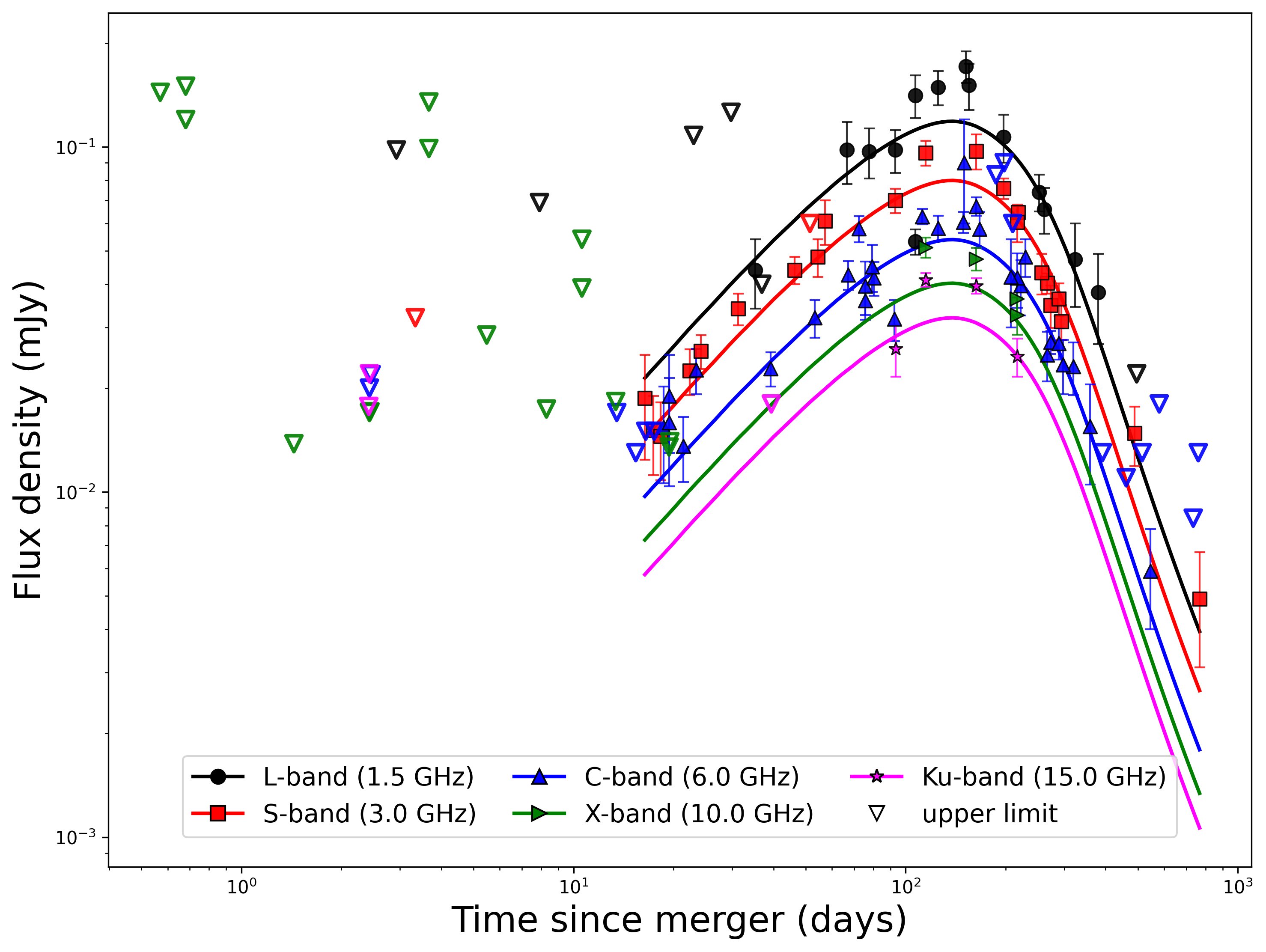}
    \caption{Federated radio light curve results for the GW170817 case study. This fit is obtained by using the GW constraints (68\% \texttt{Dingo-BNS} posterior) on the $D_L$ and $\theta_{\rm obs}$ parameters of \texttt{afterglowpy} (see \autoref{tab:AfterglowpyParams}) and a federated fitting strategy simulating eight geographically distributed private sites contributing radio observations of GW170817 (see \autoref{tab:sites}). For completeness, we also show the upper limits on the radio flux density of GW170817, corresponding to non detections, as downward-pointing triangles. These upper limits have been extracted and aggregated from GCNs using the AI-powered GCN parser developed as part of this study. They are not considered for the fit.} 
\label{fig:GW170817_lightcurve}
\end{figure}
\begin{figure*}
    \centering
    \makebox[\textwidth][c]{
        \includegraphics[width=\textwidth]{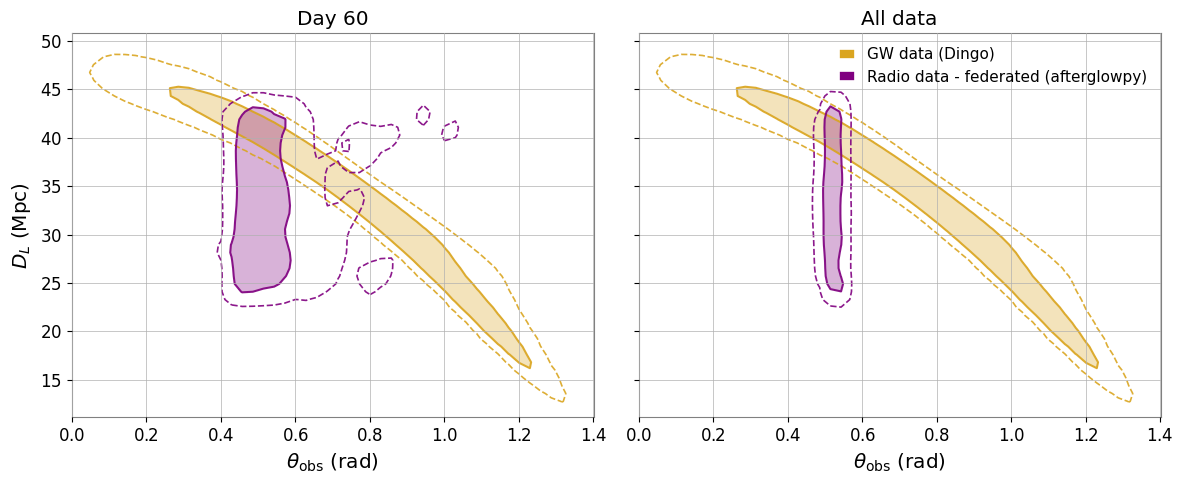}
    }
    \caption{GW170817 Case Study: Posterior distributions for the observing angle ($\theta_{\rm obs}$) and luminosity distance ($D_L$) derived using \texttt{Dingo-BNS} and our federated radio fitting approach (\autoref{sec:octopus}) are compared in the left and right panels. Contours indicate 68\% (solid) and 95\% (dashed) confidence intervals. These results are obtained by using the \texttt{Dingo-BNS}-derived GW 68\% posteriors as priors for the radio data modeled using \texttt{afterglowpy}. The left panel incorporates in the radio fit all observations collected up to 60 days post merger. The right panel incorporates all radio follow-up observations of GW170817 presented in \citet{2021ApJ...922..154M}. See  \autoref{sec:performance} for discussion.}
\label{fig:posteriors_side_by_side}
\end{figure*}
Then, the module performs an overlap analysis by comparing, for the parameters that can be constrained by both the GW and EM data, the posterior distributions from \texttt{Dingo-BNS} and that are derived as output of the radio module. 
Example results of the \texttt{Dingo-BNS} analysis for GW170817 are presented in \autoref{fig:GW170817_lightcurve} and \autoref{fig:posteriors_side_by_side}. We describe in more detail how the GW and radio posteriors are aggregated by the MMA module in \autoref{sec:octopus}. In \autoref{sec:performance}, we describe the results of this aggregation for the case of GW170817.

We stress that Bayesian inference based on the combination of GW and EM data is not new, and the $H_0$-$D_L$ inference case presented here is just one possible example of its applications. However, \texttt{RADAR}'s novelty lies the federated orchestration of Bayesian inference, allowing model-driven analysis across private/public, geographically distributed datasets without raw data sharing. 

\section{Event-driven Framework} \label{sec:octopus}
A major goal of this work is to establish a seamless connection between the GW, radio, and MMA modules, as well as a strategy for federated inference within the radio module itself. Here we describe the system that we have developed for that purpose.

\subsection{Privacy-enhancing real-time coordination tools}
Coordination and communication among the various modules described in Sections \ref{sec:gw-module}--\ref{sec:mma-module} is achieved by leveraging \texttt{Octopus} \citep{pan2024octopus}, a cloud-hosted streaming platform for developing event-driven applications and facilitating real-time communication. \texttt{Octopus} offers a managed Apache Kafka service, facilitating a hybrid cloud-to-edge event fabric to deliver information streams efficiently across distributed computational resources and enable reliable event processing at multiple sites. \texttt{Octopus} leverages Globus Auth~\cite{tuecke2016globus} to provide fine-grained authorization and access control for its highly available and fault tolerant producer and consumer interface that allows secure, low-latency event processing---a crucial feature for our MMA workflow.

While \texttt{Octopus} provides the event backbone, we leverage \texttt{ProxyStore} \citep{pauloski2023accelerating} to optimize data exchange in our federated applications via direct point-to-point data communication of large datasets such as embeddings and posterior samples. 
Instead of transferring the raw data via Octopus, \texttt{ProxyStore} allows us to pass compact proxies, deferring the data resolution to when and where the data is actually needed. This mechanism enables seamless integration with event-driven frameworks like \texttt{Octopus}, supports heterogeneous computational resources, and reduces bandwidth and memory usage---all without altering application logic. 

Together, \texttt{Octopus} and \texttt{ProxyStore} provide the foundation for privacy-enhancing, real-time coordination across geographically distributed data and analysis sites.

\subsection{Activating \sysname{}}
When a merger event is detected as output of the GW module (\autoref{sec:gw-module}), a ``Potential Merger'' event timestamp is published to the \texttt{Octopus} event fabric.  The radio module (\autoref{sec:radio-module}) continuously monitors the \texttt{Octopus} event stream and listens for LIGO-Virgo-KAGRA notices and GCNs. When a LIGO-Virgo-KAGRA alert for a ``Superevent'' (a high-confidence alert assigned a number ID in the Gravitational Wave Candidate Event Database, GraceDB) is issued, a Trigger is published in the \texttt{Octopus} event fabric if: (i) the Supervent timestamp, as reported in the LIGO-Virgo-KAGRA notice, matches the event time of a GW event identified via the GW module; 
(ii) the probability that the event is a binary neutron star merger, as estimated in the LIGO-Virgo-KAGRA notice, is $\ge$ 50\%. 

The Trigger initiates the MMA module, specifically to run \texttt{Dingo-BNS} (\autoref{sec:mma-module}), and prompts the radio module to start systematically storing GCNs in dedicated directories organized by GraceID, as described in \autoref{sec:radio-module}. We note that LIGO-Virgo-KAGRA ``Retraction Notices'' are also stored, so that when retractions occur for previously identified events, the radio module can transmit appropriate notifications through \texttt{Octopus} to enable updates of event status. 

\subsection{Event-driven federated radio light curve modeling}
\label{sec:federated}
Astronomical light curve analysis often requires aggregating flux-time data across numerous observatories. However, due to privacy constraints, data ownership policies, or the presence of sensitive proprietary data, centralizing all observational measurements is often infeasible. To overcome this challenge, within our radio module we have implemented a federated analysis system that supports both public and private datasets and enables secure, privacy-enhancing light curve inference. To this end, we have developed two complementary inference strategies, both based on the light curve modeling algorithm \texttt{afterglowpy} described in \autoref{sec:radio-module}: a distributed likelihood MCMC for exact posterior estimation and a posterior averaging method based on consensus MCMC. 

Both strategies are integrated with the \texttt{Octopus} event fabric to enable seamless communication between distributed astronomical data sites. When sufficient radio data accumulates across observation sites, the central coordinator notifies participating sites that a radio modeling task is starting, broadcasting essential information, such as the initial model parameters (see \autoref{tab:AfterglowpyParams}), the event identification name, the MCMC sampler configuration (number of walkers, iterations), and the federated approach to be used for fitting the radio light curve (distributed likelihood or posterior averaging). 

In the distributed likelihood MCMC, the coordinator proposes new parameter vectors $\theta'$ and distributes them to all sites for local likelihood evaluation. Any necessary metadata for computation tracking are also distributed to the sites. Next, each data site computes the local log-likelihood value $\log p(D_i \mid \theta')$ on its private data after receiving $\theta'$.  The server aggregates local log-likelihoods and computes the posterior probability as:
    \begin{equation}
    \log p(\theta' \mid D) \propto \log \pi(\theta') + \sum_{i=1}^{S} \log p(D_i \mid \theta').
    \end{equation}
Finally, the server accepts or rejects $\theta'$ using the MCMC ensemble sampler acceptance fraction (i.e., the fraction of proposed steps that were accepted), and the above is repeated until the maximum number of iterations is reached. 
This method yields the \textit{exact posterior} that would be obtained if all data were pooled centrally. It leverages the additive property of the log-likelihood under the assumption of data independence across sites:
\begin{equation}
\log p(D \mid \theta) = \sum_{i=1}^{S} \log p(D_i \mid \theta).
\end{equation}
Thus, even though the data remains decentralized, inference accuracy is not compromised.

The posterior averaging method is inspired by the consensus MCMC approach proposed by \citet{Scott2016,Scott2017}. The goal of this method is to improve computational efficiency to approximate the global posterior using local posterior samples computed independently at each site. Specifically, the central coordinator sends an initial guess of the model parameters $\theta^{(0)}$ to each participating data site. Then, each site executes local MCMC sampling, producing local posterior samples, $\{\theta_i^{(j)}\}_{j=1}^{N}$, where $N$ represents the number of MCMC samples. Each site then publishes \texttt{ProxyStore} references to its posterior samples via \texttt{Octopus}. These references allow the coordinator to access the samples for posterior fusion efficiently without requiring sites to transmit large data arrays directly. In the final step, the coordinator performs an aggregation step by extracting posterior samples from all proxies and approximating the global posterior using weighted average fusion. The combined posterior is computed through a weighted averaging process, where sub-posterior samples are aggregated using inverse covariances as weights. Specifically, assuming each sub-posterior is approximately Gaussian with mean \(\mu_i\) and covariance \(\Sigma_i\), the global posterior is given by:
\begin{equation}
\tilde{\Sigma}^{-1} = \sum_{i=1}^{K} \Sigma_i^{-1}, \quad 
\tilde{\mu} = \tilde{\Sigma} \sum_{i=1}^{K} \Sigma_i^{-1} \mu_i,
\end{equation}
which corresponds to Equation (2) in \citet{Scott2016}. This precision-weighted fusion approximates the product of Gaussians and yields a consensus posterior under Gaussian assumptions.

We finally note that our framework is flexible enough for the user to implement other fusion strategies such as kernel density estimation, Sequential Monte Carlo, or Mixture modeling. 

\subsection{Broadcasting results}
Upon completion of the GW parameter estimation step with \texttt{Dingo-BNS}, the MMA module publishes a ProxyStore reference containing the posterior samples to the \texttt{Octopus} event fabric. The GW parameters estimation results can be used as input of the radio module, for the purpose of establishing priors on common parameters. In this work, following \citet{2019NatAs...3..940H}, we use the \texttt{Dingo-BNS} posteriors on $D_L$ as priors for the radio light curve fitting (see \autoref{tab:AfterglowpyParams} and the yellow box in the MMA module of \autoref{fig:workflow}). 

Upon completion, the radio module publishes the posteriors derived from the federated light curve fitting via a ProxyStore reference to its posterior samples to the \texttt{Octopus} event fabric (see the purple box in the Radio module of \autoref{fig:workflow}).  

The MMA analysis is triggered once posterior samples from both \texttt{Dingo-BNS} and the radio module become available. Specifically, the availability of both posteriors enables an overlap analysis by comparing the posterior distributions of common parameters to assess the consistency between the GW signal and the associated radio counterpart (see \autoref{fig:posteriors_side_by_side}).

We stress that this federated modeling and inference architecture ensures that each site can participate in collaborative scientific inference without sharing its proprietary data. \sysname{} preserves data locality, minimizes communication overhead, and supports progressive inference, allowing estimates to improve dynamically as new observations arrive from any participating facility. We show how this works in practice for the case of GW170817 in \autoref{sec:performance}.

\begin{table*}
    \begin{center}
    \caption{Simulated private and distributed data sites for the GW170817 case study. See \autoref{sec:performance} for more details. \label{tab:sites}}
    \begin{tabular}{ccl}
    \hline 
    \hline
    Site name & Total detections & Data source label from Table 2 of \citet{2021ApJ...922..154M} \\
    \hline
    0 & 20 & \citet{alexander2018decline}, \citet{hajela2019two}, and/or \citet{margutti2018binary}\\
    1 & 0 & \citet{broderick2020lofar} \\
    2 & 1 & \citet{2019Sci...363..968G} \\
    3 & 40 & \citet{dobie2018turnover}, \citet{2017Sci...358.1579H}, \citet{2018Natur.561..355M},  \\ 
    {} & {} & \citet{2018ApJ...868L..11M},  \citet{2018Natur.554..207M}\\
    4 & 0 & \citet{kim2017alma} \\
    5 &7 &  \citet{resmi2018low} \\
    6 & 2 & \citet{Troja_2020} \\
    7 & 9 & ``This work''\\
    \hline
    \end{tabular}
    \end{center}
\end{table*}

\section{Performance Test: GW170817} \label{sec:performance}

We performed an end-to-end test of \sysname{}, using GW170817 as a case study. In what follows, we summarize the results of this test.

Running the GW module (\autoref{sec:gw-module}) on LIGO open data collected between UTC 2017-08-17 12:06:56 and 2017-08-17 13:15:12, we detect GW170817, with a combined signal-to-noise ratio of 32.4 as reported by the LIGO-Virgo Collaboration \citep{PhysRevLett.119.161101}. No false positives were recorded in the segment. To reduce the influence of a known glitch 1.1\,s before merger, we applied a Tukey window around it; the detection remains robust with or without this step. The detected ``Potential Merger'' event is assigned a timestamp of UTC 2017-08-17 12:41:03.680 by our AI models (\autoref{fig:workflow}, left box). This timestamp is then matched with the GCN issued by the LIGO-Virgo collaboration to announce the GW170817 event \citep{GCN21205}, and the radio analysis is triggered (\autoref{fig:workflow}, right box). 

The AI GCN parser is used to collect radio detections and upper limits from all public GCN notices issued for GW170817\footnote{\url{https://gcn.gsfc.nasa.gov/other/G298048.gcn3}}. The output of the AI-powered GCN parser is shown in \autoref{tab:upperlimits_table}. In addition to extracting and aggregating key information from GCNs such as GCN number, observation time, candidate EM counterpart name, observing frequency, flux density, and sky position, the GCN parser also outputs a variety of flags 
that are raised to guide the subsequent use of the extracted data for the purpose of the light curve fitting and MMA analysis. For example, GCNs  reporting upper limits (i.e., non detections) will naturally lack information such as sky position or EM candidate name,  that would be present in a GCN reporting a detection. Hence, flags are raised during the LLM interpretation to warn the user about information genuinely missing from GCNs. We raise a flag whenever an input GCN lacks: (i) frequency or flux density information, (ii) observing time; (iii) target name; (iv) sky position information. The flags are reported in the last column of Table~\ref{tab:upperlimits_table}. 

The output of the AI GCN parser is a .csv table (Table~\ref{tab:upperlimits_table}) that is passed as input to the federated fitting routine (\autoref{sec:octopus}). 
We take special consideration of the fact that multiple optical counterparts can be identified in the large GW localization errors. Hence, once the .csv table is complied, we allow the the user to select an angular radius and a sky position so that only data corresponding to sources located within a circular region of the specified angular radius centered on the specified sky coordinates are used for the radio modeling.

For this case study, we simulate a public-private federated strategy where radio follow-up data are not only aggregated from public GCNs as described above, but also collected from a suite of private and geographically distributed sites that wish to preserve data rights. To this end, we divided the GW170817 radio dataset presented in \citet{2021ApJ...922..154M} into eight different private sites (see \autoref{tab:sites} for details).

As described in Sections~\ref{sec:methods} and \ref{sec:octopus}, a federated fitting strategy is then orchestrated where  \texttt{Dingo-BNS} returns 68\% confidence level posteriors for the luminosity distance $D_L$ and the observing angle $\theta_{\rm obs}$ derived using open GW data from the LIGO detectors. The \texttt{Dingo-BNS}  posteriors are then used as priors for a federated radio fit of the same parameters based on \texttt{afterglowpy}, and on data aggregated from public GCNs and private sites (Tables~\ref{tab:sites} and~\ref{tab:upperlimits_table}). The posterior distributions for $\theta_{\rm obs}$ and $D_L$ derived using \texttt{Dingo-BNS} and our federated radio fitting approach (\autoref{sec:octopus}) are shown in \autoref{fig:posteriors_side_by_side}. Contours in this figure indicate the 68\% (solid) and 95\% (dashed) confidence intervals. The left panel presents results obtained by combining \texttt{Dingo-BNS}-derived GW posteriors with radio data modeled using \texttt{afterglowpy}, incorporating observations collected up to 60 days post merger. Our approach supports progressive inference, allowing estimates to improve dynamically as new observations arrive from any participating 
facility or observer. Indeed, in the right panel of \autoref{fig:posteriors_side_by_side} we show how the results are updated once the full radio dataset compiled in \citet{2021ApJ...922..154M} is made available. The constraints on both $D_L$ and $\theta_{\rm obs}$ improve significantly as more radio data are included. The overlay of GW posteriors (yellow) with the evolving radio constraints (purple) highlights the value of early MMA analyses in guiding follow up strategies. For example, the day-60 result (left panel in \autoref{fig:posteriors_side_by_side}) strongly indicates an off-axis geometry which, given that larger $\theta_{\rm obs}$ correspond to later-peaking radio afterglows, would motivate continued radio monitoring. As shown in the right panel of \autoref{fig:posteriors_side_by_side}, observations extending through the light curve peak ($\approx 200$\,days post-merger; see \autoref{fig:GW170817_lightcurve}) yield tighter constraints on $\theta_{\rm obs}$, with implications on physical inferences such as $H_0$ estimation \citep{2019NatAs...3..940H}.

\section{Resilience} \label{sec:resilience}
A key consideration in the adoption of scalable frameworks, such as \sysname{}, is their resilience to potential failures. While a comprehensive evaluation of failure modes and corresponding stress tests is beyond the scope of this work, we offer preliminary insights by examining two critical components of our framework. First, we assess the system’s ability to perform federated data analysis despite communication disruptions with geographically distributed, privacy-sensitive data contributors. Second, we evaluate the robustness of the AI-powered GCN interpreter, and particularly the performance of the underlying language model, in reliably extracting structured information from unformatted public GCN circulars. These components are central to \sysname{}, as the quality and reliability of the resulting MMA analysis, which may ultimately influence follow-up strategies and telescope scheduling, depend directly on both private and public data inputs.

\begin{table}[htbp!]
    \begin{center}
    \caption{Performance metrics of the resilience of the AI-powered GCN parser, which integrates \texttt{GPT-4.1} as its language modeling component. Metrics include precision, recall, and F1 score for event matching, as well as GCN-specific recall (GCN-R), reported as mean values with standard deviations across test conditions.
    \label{tab:GCNparser_metric}
    }
    \begin{tabular}{lcc}
    \hline 
    \hline
    Metric & Mean & Standard Deviation \\
    \hline
    Event match precision, $P$ & 0.819 & 0.023 \\
    Event match recall, $R$ & 0.727 & 0.047 \\
    Event match F1 & 0.770 & 0.035 \\
    GCN match recall, GCN-R & 0.724 & 0.033 \\
    \hline
    \end{tabular}
    \end{center}
\end{table}

For the first resilience study, 
we simulate communication interruptions across multiple sites (see \autoref{tab:sites}) by testing the following four scenarios: 
\begin{enumerate}
\item Removal of the four sites contributing the least data (sites 1, 2, 4 and 6 in \autoref{tab:sites}); 
\item Retention of only the two highest-contributing sites (sites 0 and 3 in \autoref{tab:sites}); 
\item Exclusion of the most data-rich site (site 3 in \autoref{tab:sites}); 
\item Omission of key light curve segments near the peak and during the interval between 60 and 200 days post-merger (see \autoref{fig:GW170817_lightcurve}). 
\end{enumerate}

\begin{figure}[htbp!]
    \centering
\includegraphics[width=\columnwidth]{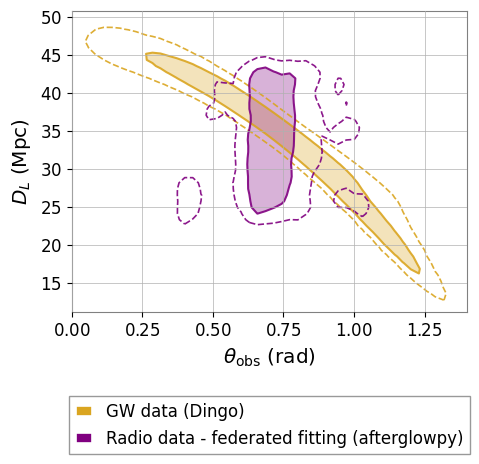}
    \caption{Posterior distributions for the observing 
    angle ($\theta_{\rm obs}$) and luminosity distance ($D_L$) 
    derived using \texttt{Dingo-BNS} and our federated radio fitting 
    approach, when, as per resilience scenario (3) in the text, 
    he site with the most data points (site 3 in \autoref{tab:sites}) is omitted. The procedure for 
    federated fitting used here is as in 
    \autoref{fig:posteriors_side_by_side}.
    } 
\label{fig:GW170817_resilience_test}
\end{figure}

We observe that \sysname{} completed without errors in all four cases: i.e., it ran to completion, producing a federated fit as designed.
We assessed the resulting federated fits and determined that they appropriately reflected the absence of critical (broadening in the MMA posterior contours) or non-critical (little effect on posterior contours) data. As an example, me show the result for scenario (3) in \autoref{fig:GW170817_resilience_test}. 

In the reported failure scenarios, we simulated site dropouts by excluding 
data contributors at the start of execution. As such, latency remained 
consistent with the failure-free case (3--4 hours), as the federated fitting 
proceeded without waiting on missing nodes. We found that runtime is dominated 
by the number of MCMC iterations; thus, introducing convergence-based early 
stopping would offer more substantial gains. In a real-time deployment, 
dynamic failures (e.g., mid-run site dropouts or high-latency responses) would indeed increase latency due to blocking communication. To mitigate this, 
we plan to implement timeout mechanisms that allow the system to proceed 
with partial site participation, preserving responsiveness for time-critical 
MMA follow-up.

In our second resiliency analysis, we assess the reliability 
of our AI-powered GCN interpreter, which leverages 
\texttt{GPT-4.1}, 
in extracting structured data from unformatted, free-text 
GCN circulars. These circulars vary widely in format, 
terminology, and clarity, presenting a significant challenge 
for automated parsing. By comparing multiple independent model 
outputs against expert-curated ground truth and quantifying 
extraction accuracy across several metrics, we probe the 
interpreter's ability to consistently deliver usable 
information despite inherent variability in its inputs. 
Specifically, we analyze six independent sets of prediction 
outputs generated for the complete set of GW170817-related 
GCN circulars, and compare them to ground truth values.

We obtained the ground truth, defined as columns 1--7 in  \autoref{tab:upperlimits_table}, by having a trained expert manually extract and curate the information contained in the GCNs. To quantify performance, we define two levels of matching. Event-level matching requires that each individual observation parsed from a GCN exactly reproduce the expert-assigned values for columns 1--7 in \autoref{tab:upperlimits_table}. GCN-level matching extends this criterion to the full set of events within a single GCN, such that the GCN is considered a match only if all of its observations meet the event-level criterion. To measure these matches, we define the following metrics:

\begin{itemize}
\item Event match precision, $P$ = \# matched AI-parser events / \# total events outputted by the AI parser;
\item Event match recall, $R$ = \# matched AI-parser events / \# total ground truth events;
\item Event match F1 score = $2 \times (P\times R) / (P + R)$;
\item 
GCN match recall, GCN-R = \# GCNs matched by the AI-parser  / \# total GCNs.
\end{itemize}

We noted a clear difference in the reliability of the AI interpreter depending on whether a GCN reports a detection or an upper limit. Specifically, in the case of GCNs reporting detections (see the two lines marked with asterisks in Table \ref{tab:upperlimits_table}), the output of the AI interpreter is consistently in agreement with the ground truth. Hence, all of the above metrics are maximized. On the other hand, the metrics for GCNs reporting upper limits drop as shown in \autoref{tab:GCNparser_metric} (additional resilience results 
for an AI-powered GCN parser that integrates 
\texttt{Llama3.1-405B-Instruct} are provided 
in~\autoref{tab:GCNparser_metric_llama}). The reduced values of these metrics in the case of GCNs reporting upper limits reveal the challenges posed by the unstructured, free-text format of human-readable circulars used during the GW170817 follow-up campaign. Enhanced standardization such as that of GCN notices and, more specifically, the inclusion of structured fields for key observational parameters like time, frequency, and flux density, would significantly improve the accuracy and reliability of automated data extraction and aggregation from GCNs. 

Overall, this analysis highlights both the strengths and limitations of using LLMs for scientific data extraction under real-world conditions, framing model reliability not just in terms of accuracy, but in its robustness to semantic ambiguity and structural inconsistency. We stress, however, that because only detections are used for the MMA analysis performed by \texttt{RADAR}, the above noted inaccuracies do not impact downstream tasks and if we were to run \texttt{RADAR} twice, once with the GCN data extracted by a trained the expert and once with the data extracted by the AI-powered GCN interpreter, there would be no difference in the output. Finally, we briefly discuss in \autoref{sec:appendix_model} how different LLMs performed in this study.

\section{Summary and conclusion}
\label{sec:conclusion}

We have presented \sysname{}, a framework aimed at addressing key challenges in the EM follow-up of GW events. Critical among these challenges are the large GW localization regions, which contrast sharply with the comparatively narrow fields of view of sensitive radio arrays, and the requirement for prolonged radio monitoring extending years post-merger to derive accurate astrophysical constraints on the emitting sources and their properties. The demanding observing resource requirements imposed by these two challenges can be mitigated through enhanced global coordination among observers and telescopes. \sysname{} seeks to advance this objective by enabling community-driven information sharing, federated data analysis, and system resilience, while incorporating AI techniques for both GW signal identification and radio data aggregation and analysis. We note that RADAR can also be employed to facilitate follow up observations and modeling efforts within multiple networks of collaborations, not necessarily as a single centralized effort.

Using GW170817 as a case study, we have shown that it is possible to maintain data ownership and privacy while enabling the sharing of predictive models that can guide and optimize follow-up strategies. Although our focus has been on joint GW and radio analyses, the framework presented here is flexible and could be readily extended to a variety of triggers identified in other bands of the EM spectrum \citep[e.g.,][]{2021MNRAS.501.3272M,2025A&A...698A.153S,2025A&A...694A.183B}, as well as events in other bands of the GW spectrum \citep[e.g.,][and references therein]{2021ApJ...911L..15Y,2022PhRvD.106j3017M,2023LRR....26....2A,2024MNRAS.533.3164X,2025ApJ...979..155P}. It can also be adapted to incorporate diverse model predictions tailored to different users and scientific objectives. While we have begun to explore the resilience of the framework, a more comprehensive evaluation of potential failure points will be essential to ensure its broader applicability. We envision this to become possible during future observing runs of ground-based GW detectors, such as the LIGO-Virgo-KAGRA O5 \citep{2018LRR....21....3A}, as binary neutron star mergers localized to $\lesssim 100$\,deg$^2$ should come into reach and optical facilities such as Rubin could enable the identification (and more accurate localization) of their associated kilonovae \citep{{2022ApJS..260...18A}}. This will open the way to a hunt for radio afterglows similar to that of GW170817, something the community has been waiting for since the O2 run.

In the longer-term future, when improved GW detectors may issue pre-merger alerts providing GW localization areas bettern than 100\,deg$^2$ at least 5\,min before merger \citep[see e.g., Figure  in 6][]{2024FrASS..1186748C}, an important evaluation metric for AI-driven multi-messenger frameworks such as \texttt{RADAR} will become the latency with which data from heterogeneous observatories are acquired, processed, and fused to yield scientifically actionable outputs. This metric is critical for enabling rapid follow-up observations, including the tantalizing possibility of pre-merger alerts that could capture prompt emission from binary neutron star mergers—phenomena that remain largely unexplored \citep[e.g.,][and references therein]{10.1093/mnras/stw576}.
In the GW170817 case study, \texttt{RADAR} demonstrated an end-to-end latency on the order of a few hours. However, this aggregate timescale masks the varying latencies of individual pipeline components, each operating within expected bounds. The GW module exemplifies low-latency performance: it processed 4096\,s of data from LIGO Hanford and Livingston, sampled at 4096\,Hz, in under 4.5 minutes using a batch size of 8\,s. The GCN parser completed its task in about 4 minutes, while the \texttt{Dingo-BNS} module required less than 10 minutes to download relevant data and infer $(\theta_{\rm obs}, D_L)$. Communication across modules via \texttt{Octopus} typically took less than 2--3 seconds.
By contrast, the dominant contributor to overall latency was the modeling of the radio data, which involves federated, multi-site fitting of slowly evolving afterglow light curves. Using a distributed log-likelihood MCMC approach, each of 32 walkers proposes a $(\theta_{\rm obs}, D_L)$ pair. These are dispatched to eight geographically distributed radio data sites via \texttt{Octopus}. Each site evaluates its local log-likelihood given the proposed parameters, and the collective results are aggregated to compute the global posterior. This process, which takes approximately 3--4 hours, reflects the inherent timescale of analyzing radio transients that evolve over days to years. Notably, even in a non-federated, centralized setting, this latency would persist due to the cadence and nature of radio observations, which often require $\sim$1 hour per integration.

\texttt{RADAR}'s key innovation lies in its ability to coherently orchestrate multi-messenger inference across disparate data modalities with minimal overhead. This capability is not only essential for time-domain astrophysics today but will become increasingly critical as next-generation detectors boost sensitivity and event rates \citep[e.g.,][]{Nitz:2021pbr}.

In future work, we plan to systematically analyze the factors 
contributing to end-to-end latency, benchmark \texttt{RADAR}'s 
performance against alternative frameworks, and 
explore algorithmic and infrastructural strategies 
to further reduce response times.

\begin{acknowledgments}
We thank Stephen Green, Maximilian Dax, and Michael Pürrer for their assistance with the use of \texttt{Dingo-BNS}. A.C. and K.M. acknowledge partial support from the National Science Foundation (NSF) via grant \# AST-2431072 and from the U.S. Department of Energy (DoE) via grant \# DE-SC0025935. E.A.H. acknowledges support from NSF grants OAC-2514142 and OAC-2209892.  This work was partially carried out at the Advanced Research Computing at Hopkins (ARCH) core facility  (rockfish.jhu.edu). This research has made use of data obtained from the Gravitational Wave Open Science Center (\url{https://gwosc.org}), a service of the LIGO Scientific Collaboration, the Virgo Collaboration, and KAGRA. This work was partially supported by the U.S. Department of Energy under Contract No. DE-AC02-06CH11357, including funding from the Office of Advanced Scientific Computing Research (ASCR)'s Diaspora project and the Laboratory Directed Research and Development program.  An award for computer time  was provided by the U.S. Department of Energy’s  
Innovative and Novel Computational Impact on Theory and 
Experiment (INCITE) Program. This research used supporting 
resources at the Argonne and the Oak Ridge Leadership 
Computing Facilities. The Argonne Leadership Computing 
Facility at Argonne National Laboratory is supported 
by the Office of Science of the U.S. DOE under Contract 
No. DE-AC02-06CH11357. The Oak Ridge Leadership 
Computing Facility at the Oak Ridge National Laboratory 
is supported by the Office of Science of the U.S. DOE 
under Contract No. DE-AC05-00OR22725. This research used 
both the DeltaAI advanced computing and data resource, 
which is supported by the National Science Foundation 
(award OAC 2320345) and the State of Illinois, and the 
Delta advanced computing and data resource which is 
supported by the National Science Foundation 
(award OAC 2005572) and the State of Illinois. 
Delta and DeltaAI are joint efforts of the 
University of Illinois Urbana-Champaign and 
its National Center for Supercomputing Applications. The scientific software used to reproduce and extend the results presented in this manuscript is available in the \texttt{GitHub} repository~\citet{patel2025ai4mma_code}.
\end{acknowledgments}

\appendix

\section{AI model for gravitational wave detection}
\label{sec:appendix_model}

This appendix outlines the architecture used in our gravitational wave detection model. While the high-level structure is inspired by the model presented in \citet{tian2023aiensemble,Tian:2023vdc}, our implementation introduces a key change in how detector outputs are aggregated. Specifically, we replace the graph-based aggregation mechanism with a cross-channel communication scheme that uses learned pairwise projections between detectors via a multi-head attention mechanism \citep{vaswani2017attention}. This modification enables more expressive modeling of detector-specific correlations and has contributed significantly to reducing false positives. The full architecture consists of three stages:

\begin{itemize}
    \item \textit{Hierarchical Dilated Convolutional Network (HDCN):} Captures multi-scale temporal features within each individual detector's strain data using dilated convolutions and gated activations.
    \item \textit{Cross-Attention Network (CAN):} Exchanges information between detectors using multi-head attention-based projections to dynamically align and weight channel-wise features.
    \item \textit{Final Output Module:} Combines the updated detector representations to produce a unified per-timestep prediction score.
\end{itemize}

We refer here to the combination of the cross-attention network (CAN) and the final output module as the \textit{aggregator}, which replaces the graph-based aggregation used in~\citet{Tian:2023vdc}.

\subsection{Hierarchical Dilated Convolutional Network (HDCN)}

\begin{figure}[htbp]
\centering
\includegraphics[width=0.48\textwidth]{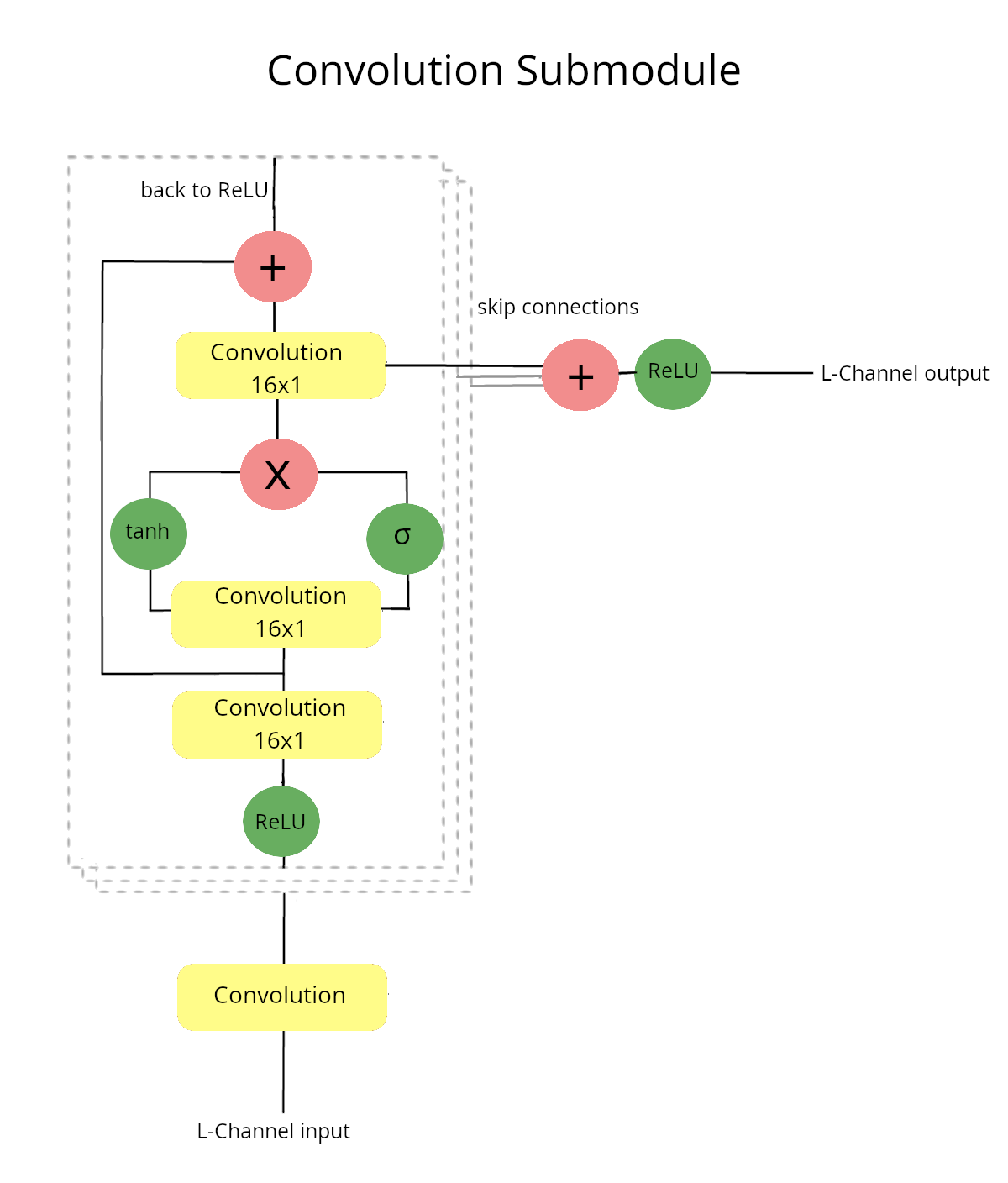}
\caption{Hierarchical Dilated Convolutional Network (HDCN) submodule for the Livingston  strain channel. An identical module is used for the Hanford strain channel. Each block consists of dilated convolutions with skip connections and nonlinear gating.
}
\label{fig:hdcn_blocks}
\end{figure}

The HDCN consists of $N=33$ layers with exponentially increasing dilation 
factors ($d = 1, 2, 4, \dots, 1024$), repeated over three cycles. This 
configuration yields a receptive field of approximately $T = 1.5$ seconds 
(given a sampling rate of 4096 Hz), sufficient to encompass the full 
inspiral-merger-ringdown sequence for binary neutron star events. Each strain channel (LIGO Hanford or LIGO Livingston) is processed independently through a stack of dilated 1D convolutional layers. These layers use exponentially increasing dilation rates to capture both short-term and long-range dependencies in the signal while maintaining the original sequence resolution. Each convolutional unit includes gated activation paths (\texttt{tanh} and \texttt{sigmoid}), skip connections, and ReLU activations to stabilize training and promote feature reuse. This architecture component is schematically 
shown in~\autoref{fig:hdcn_blocks}.

\subsection{Cross-Attention Network (CAN)}

\begin{figure}[htbp]
\centering
\includegraphics[width=0.75\textwidth]{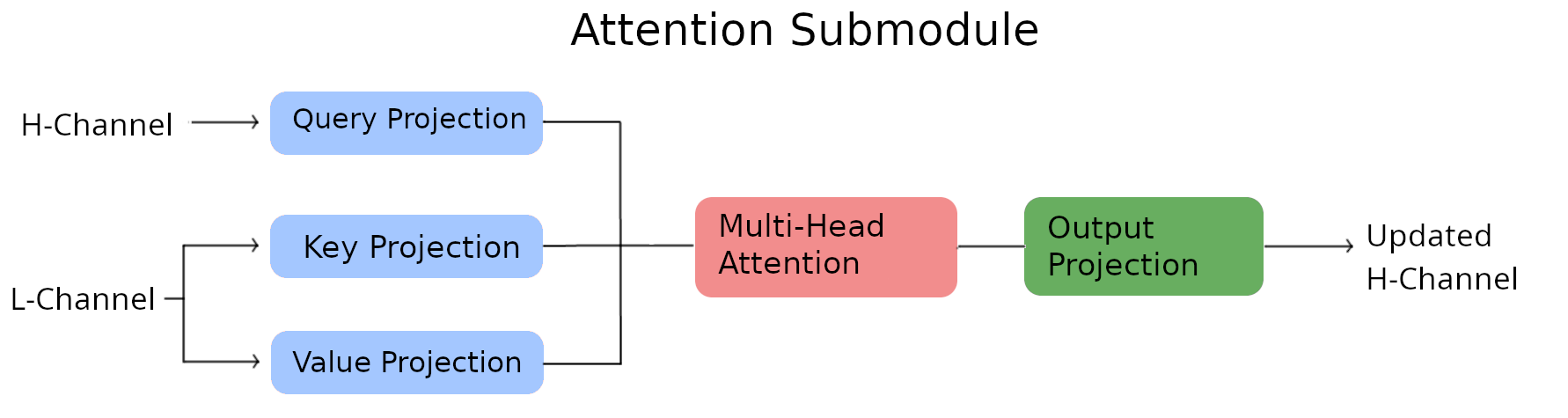}
\includegraphics[width=0.75\textwidth]{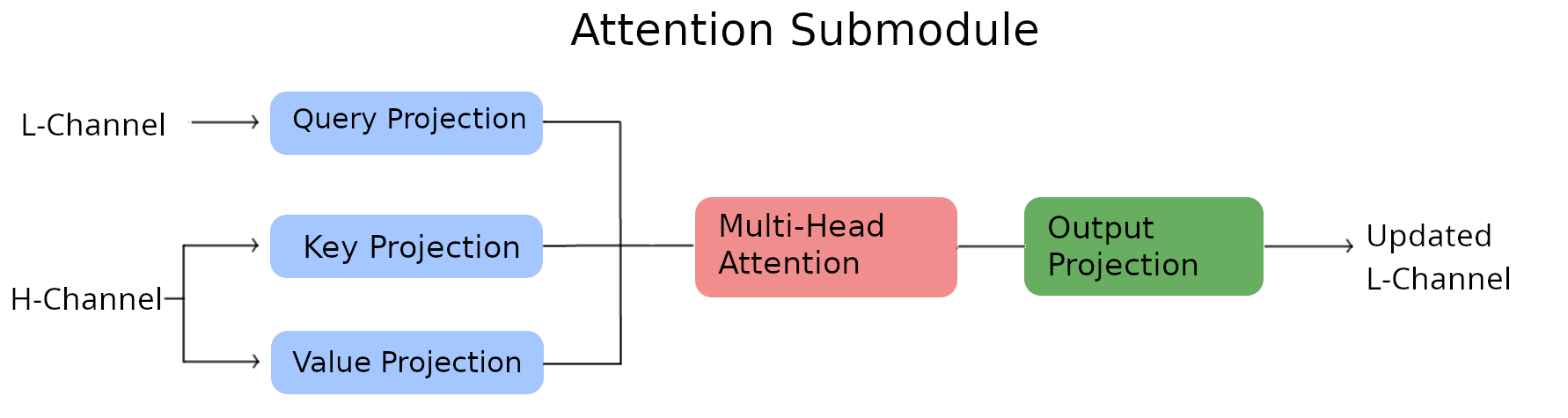}
\caption{Cross-Attention Network (CAN) submodules. Top: Hanford features attending to Livingston; Bottom: Livingston features attending to Hanford. Each module projects features to query, key, and value spaces before performing multi-head attention and output projection.}
\label{fig:can_blocks}
\end{figure}

Following local feature extraction, each detector channel communicates with the other through a Cross-Attention Network (CAN). In this step, features from one detector serve as the \textit{query}, while those from the other serve as \textit{key} and \textit{value} inputs in a multi-head attention scheme. This architecture allows the model to dynamically weight and align information across detectors on a per-timestep basis. The output is an updated sequence for each detector, enriched with information from its counterpart. This architecture component is schematically 
shown in~\autoref{fig:can_blocks}.

\subsection{Final Output Module}

\begin{figure}[htbp]
\centering
\includegraphics[width=0.75\textwidth]{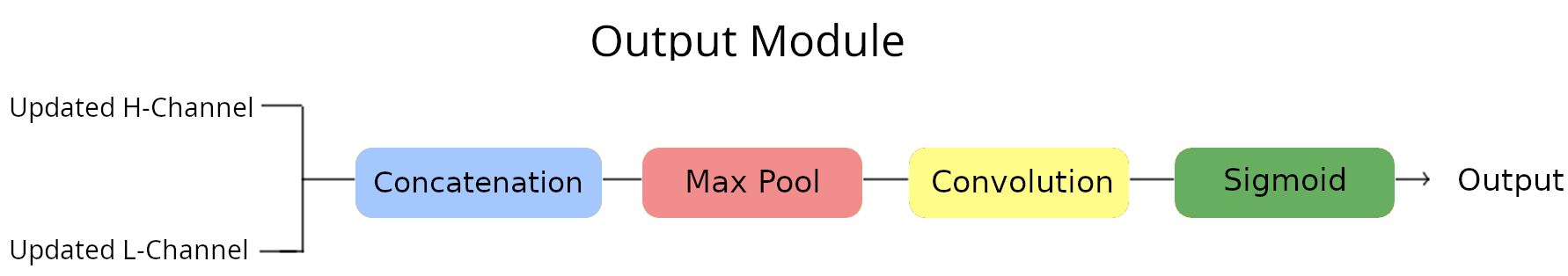}
\caption{Final Output Module. The enriched features from both detectors are concatenated, pooled, and projected to generate the per-timestep output score.}
\label{fig:output_module}
\end{figure}

The final module combines the updated detector representations and computes the model’s prediction. Features from Hanford and Livingston are concatenated along the channel dimension and passed through a max-pooling operation, a 1D convolution, and a sigmoid activation. This produces a score at each timestep, indicating the model's confidence that a gravitational wave signal is present in the corresponding segment. The structure of the final output module is shown in~\autoref{fig:output_module}.

\section{AI-powered GCN parser}
To rigorously evaluate the performance of our AI-powered GCN parser for extracting structured data from unstructured GCN circulars, we adopt a zero-shot prompting strategy, in which the LLM is instructed to extract all reported radio follow-up events from a given GCN notice and return them in a strict JSON format. Each predicted event is considered correctly matched only if all required fields—GCN number, observation time, source name, observing frequency, flux density, right ascension (RA), and declination (DEC)—are correctly extracted. We assess the parser's performance using four metrics: (1) Event match precision, defined as the ratio of correctly matched LLM-predicted events to the total number of LLM-predicted events; (2) Event match recall, defined as the ratio of correctly matched LLM-predicted events to the total number of ground-truth events; (3) Event match F1 score, computed as the harmonic mean of precision and recall; and (4) GCN match recall, which measures the fraction of GCN circulars for which all reported events are perfectly matched in the LLM's predictions. In cases where required fields are missing from the output, the event is marked as unmatched. If the LLM response does not adhere to the expected JSON format, we implement an automated re-prompting mechanism with up to three retries. Should all retries fail, the parser is deemed unable to process that particular GCN, and the result is recorded as a failure for that instance. This evaluation framework ensures robust and transparent assessment of the LLM's ability to convert heterogeneous, human-authored GCN reports into structured, machine-readable formats suitable for downstream analysis.

The information pertaining to the light curves is extracted from the GCNs using a series of prompts passed to \texttt{GPT-4.1} 
which aim at returning the radio data in the most consistent and predictable way. A system prompt is used when defining the parser in order format and to ensure consistency in the outputs. A series of three prompts are then passed in to the LLM. These are, in summary:

\begin{enumerate}
    \item Does this GCN describe a radio follow-up to an electromagnetic counterpart? (yes/no).
    \item Does this GCN include an explicit statement of non-detection?
    \item Please return the properties of observation (time since GW, observation frequency, flux density, etc.).
\end{enumerate}
The responses are used to create \autoref{tab:upperlimits_table}, which also contains the ground truths. 

\setlength{\tabcolsep}{3pt}
\setlength\LTcapwidth{\textwidth}
\begin{longtable*}{cccccccc}
\caption{Summary of the observations related to GW170817 as reported in the GCN circulars. Column~1 lists the GCN circular number. Column~2 provides either the observation time (when explicitly stated) or the GCN publication timestamp. Column~3 records the name of the candidate electromagnetic (EM) counterpart, when available. Column~4 indicates the observing frequency. Column~5 reports the measured flux density or, in the absence of a detection, the upper limit on the flux density. Columns~6 and 7 give the sky coordinates (Right Ascension and Declination in hh:mm:ss and dd:mm:ss, respectively) of the candidate EM counterpart, when provided. Column~8 shows the number of flags (see \autoref{sec:performance} for a detailed description); `NA' marks entries for which \texttt{GPT 4.1} never returned the ground truth even over multiple runs. A superscript asterisk ($^*$) denotes a radio detection; all other entries represent non-detections (upper limits). The values in Columns~1--7 represent what would be produced by a trained expert through manual extraction and curation of GW170817-related GCN circulars.}
\label{tab:upperlimits_table} \\

\hline
\colhead{Number} & \colhead{Date and Time} & \colhead{Name} &
\colhead{Frequency (GHz)} & \colhead{Flux Density (mJy)} &
\colhead{R.A.} & \colhead{Dec.} & \colhead{Flags} \\
\hline
\endfirsthead

\multicolumn{8}{c}%
{{\bfseries \tablename\ \thetable{} -- continued from previous page}} \\
\hline
\colhead{Number} & \colhead{Date and Time} & \colhead{Name} &
\colhead{Frequency (GHz)} & \colhead{Flux Density (mJy)} &
\colhead{R.A.} & \colhead{Dec.} & \colhead{Flags} \\
\hline
\endhead

\hline \multicolumn{8}{r}{{Continued on next page}} \\
\endfoot

\hline
\endlastfoot
21545$^*$ & 17/08/18 02:09:00 GMT &  & 9.77 & 0.3 & 13:09:47.850 & -23:23:01.29 & 3 \\
21571 & 84/06/10 12:00:00 GMT &  & 1.51 & 0.54 &  &  & 5\\
21571 & 84/06/10 12:00:00 GMT &  & 4.89 & 1.02 &  &  & 5 \\
21571 & 84/06/23 12:00:00 GMT &  & 1.51 & 0.54 &  &  & NA \\
21571 & 84/06/23 12:00:00 GMT &  & 4.89 & 1.02 &  &  & NA \\
21574 & 17/08/18 20:19:00 GMT & SSS17a & 8.5 & 0.12 &  &  & 4\\
21574 & 17/08/18 20:19:00 GMT & SSS17a & 10.5 & 0.15 &  &  & 4 \\
21574 & 17/08/18 20:19:00 GMT & SSS17a & 16.7 & 0.13 &  &  & 4 \\
21574 & 17/08/18 20:19:00 GMT & SSS17a & 21.2 & 0.14 &  &  & 4 \\
21589 & 17/08/18 22:04:57 GMT &  & 10.0 & 0.017 & 13:09:47.704 & -23:23:02.45 & 2 \\
21613 & 17/08/19 22:01:48 GMT & SSS17a & 9.7 & 0.018 &  &  & 3 \\
21614 & 17/08/19 22:01:48 GMT & SSS17a & 6.0 & 0.022 & 13:09:48.089 & -23:22:53.35 & 2 \\
21636 & 17/08/19 22:01:48 GMT & SSS17a & 15.0 & 0.016 & 13:09:48.089 & -23:22:53.35 & 2 \\
21650 & 17/08/20 12:00:00 GMT & SSS17a & 2.0 & 0.04 &  &  & 2 \\
21664 & 17/08/22 12:00:00 GMT & SSS17a/DLT17ck & 6.0 & 0.02 &  &  & 3 \\
21670 & 17/08/20 23:31:02 GMT & SSS17a & 8.5 & 0.135 &  &  & 3 \\
21670 & 17/08/20 23:31:02 GMT & SSS17a & 10.5 & 0.099 &  &  & 3 \\
21708 & 17/08/20 11:00:00 GMT & SSS17a & 1.4 & 0.1 &  &  & NA \\
21740 & 17/08/27 23:26:24 GMT & SSS17a & 8.5 & 0.054 &  &  & 3 \\
21740 & 17/08/27 23:26:24 GMT & SSS17a & 10.5 & 0.039 &  &  & 3 \\
21750 & 17/08/19 14:55:15 GMT & SSS17a & 97.5 & 0.05 &  &  & 4 \\
21750 & 17/08/26 14:55:15 GMT & SSS17a & 97.5 & 0.05 &  &  & NA \\
21760 & 17/08/19 21:57:46 GMT & SSS17a & 13.0 & 50.0 &  &  & 2 \\
21760 & 17/08/23 21:18:44 GMT & SSS17a & 13.0 & 50.0 &  &  & 2 \\
21763 & 17/08/26 12:15:00 GMT & SSS17a & 5.0 & 0.16 &  &  & 3 \\
21768 & 17/08/25 10:30:00 GMT & SSS17a & 1.39 & 0.13 &  &  & 3 \\
21768 & 17/08/25 10:30:00 GMT & SSS17a & 1.39 & 0.16 & 13:09:47.74 & -23:23:01.24 & 2 \\
21803 & 17/09/01 12:00:00 GMT & SSS17a & 16.7 & 0.05 &  &  & 3 \\
21803 & 17/09/01 12:00:00 GMT & SSS17a & 21.2 & 0.05 &  &  & 3 \\
21803 & 17/09/01 12:00:00 GMT & SSS17a & 43.0 & 0.09 &  &  & 3 \\
21803 & 17/09/01 12:00:00 GMT & SSS17a & 45.0 & 0.09 &  &  & 3 \\
21804 & 17/08/23 13:00:00 GMT & SSS17a & 5.0 & 0.18 &  &  & 3 \\
21804 & 17/08/24 13:00:00 GMT & SSS17a & 5.0 & 0.16 &  &  & 3 \\
21804 & 17/08/25 13:00:00 GMT & SSS17a & 5.0 & 0.16 &  &  & 3 \\
21804 & 17/08/26 13:00:00 GMT & SSS17a & 5.0 & 0.19 &  &  & 3 \\
21804 & 17/08/27 13:00:00 GMT & SSS17a & 5.0 & 0.15 &  &  & 3 \\
21804 & 17/08/28 13:00:00 GMT & SSS17a & 5.0 & 0.15 &  &  & 3 \\
21804 & 17/09/04 07:48:43 GMT & SSS17a & 5.0 & 0.07 &  &  & 4 \\
21850 & 17/08/20 12:00:00 GMT & SSS17a/DLT17ck & 8.7 & 0.125 &  &  & 4 \\
21850 & 17/08/21 12:00:00 GMT & SSS17a/DLT17ck & 8.7 & 0.12 &  &  & 4 \\
21850 & 17/09/08 11:16:27 GMT & SSS17a/DLT17ck & 8.7 & 0.088 &  &  & 4 \\
21882 & 17/09/08 12:00:00 GMT & SSS17a & 16.7 & 0.035 &  &  & 3 \\
21882 & 17/09/08 12:00:00 GMT & SSS17a & 21.2 & 0.035 &  &  & 3 \\
21891 & 17/09/06 12:00:00 GMT & SSS17a & 1.5 & 0.075 &  &  & 3 \\
21899 & 17/08/20 01:44:32 GMT & SSS17a & 1.4 & 1400.0 &  &  & NA \\
21899 & 17/08/20 02:50:14 GMT & SSS17a & 1.4 & 1400.0 &  &  & NA \\
21900 & 17/09/15 12:00:00 GMT & SSS17a & 16.7 & 0.042 &  &  & 1 \\
21900 & 17/09/15 12:00:00 GMT & SSS17a & 21.2 & 0.042 &  &  & 1 \\
21914 & 17/09/07 12:00:00 GMT & SSS17a & 7.2 & 1.5 &  &  & 3 \\
21914 & 17/09/08 12:00:00 GMT & SSS17a & 7.2 & 1.5 &  &  & 3 \\
21914 & 17/09/09 12:00:00 GMT & SSS17a & 7.2 & 1.5 &  &  & 3 \\
21914 & 17/09/19 12:00:00 GMT & SSS17a & 7.2 & 1.5 &  &  & 3 \\
21914 & 17/09/22 19:06:44 GMT & SSS17a & 7.2 & 0.5 &  &  & 4 \\
21914 & 17/09/22 19:06:44 GMT & SSS17a & 7.2 & 1.0 &  &  & NA \\
21914 & 17/09/22 19:06:44 GMT & SSS17a & 7.2 & 3.0 &  &  & NA \\
21920 & 17/09/10 13:35:00 GMT & SSS17a & 32.0 & 30.0 &  &  & 3 \\
21920 & 17/09/15 12:00:00 GMT & SSS17a & 14.6 & 0.076 &  &  & 3 \\
21920 & 17/09/15 12:00:00 GMT & SSS17a & 14.6 & 0.052 &  &  & 3 \\
21920 & 17/09/15 12:00:00 GMT & SSS17a & 14.6 & 320.0 &  &  & 3 \\
21920 & 17/09/15 12:00:00 GMT & SSS17a & 14.6 & 190.0 &  &  & 3 \\
21920 & 17/09/15 12:00:00 GMT & SSS17a & 14.6 & 150.0 &  &  & 3 \\
21920 & 17/09/15 12:00:00 GMT & SSS17a & 14.6 & 100.0 &  &  & 3 \\
21927 & 17/08/18 07:07:00 GMT &  & 0.185 & 17.0 &  &  & NA \\
21928 & 17/08/18 06:49:31 GMT & SSS17a & 1.4 & 1400.0 & & & 2 \\
21928 & 17/08/18 08:50:36 GMT & SSS17a & 1.4 & 1400.0 & &  & 2 \\
21929$^*$ & 17/09/17 12:00:00 GMT & SSS17a & 3.0 & 0.034 &  & & 3\\
21933 & 17/08/26 08:43:00 GMT & SSS17a & 1.48 & 0.07 &  &  & 3 \\
21933 & 17/09/06 03:22:00 GMT & SSS17a & 1.48 & 0.075 &  &  & 3 \\
21933 & 17/09/17 07:16:00 GMT & SSS17a & 1.48 & 0.06 &  &  & 3 \\
21939 & 17/09/20 10:00:00 GMT & SSS17a & 5.0 & 0.14 &  &  & 3 \\
21940 & 17/08/23 15:36:28 GMT & SSS17a & 5.0 & 0.11 &  &  & 3 \\
21940 & 17/08/24 15:33:42 GMT & SSS17a & 5.0 & 0.09 &  &  & 3 \\
21940 & 17/08/25 15:30:26 GMT & SSS17a & 5.0 & 0.097 &  &  & 3 \\
21940 & 17/08/26 15:30:26 GMT & SSS17a & 5.0 & 0.112 &  &  & 3 \\
21940 & 17/08/27 15:30:22 GMT & SSS17a & 5.0 & 0.087 &  &  & 3 \\
21940 & 17/08/28 15:30:26 GMT & SSS17a & 5.0 & 0.082 &  &  & 3 \\
21940 & 17/08/31 15:01:26 GMT & SSS17a & 5.0 & 0.109 &  &  & 3 \\
21940 & 17/09/01 15:01:26 GMT & SSS17a & 5.0 & 0.114 &  &  & 3 \\
21940 & 17/09/02 15:04:42 GMT & SSS17a & 5.0 & 0.144 &  &  & 3 \\
21940 & 17/09/03 15:01:26 GMT & SSS17a & 5.0 & 0.166 &  &  & 3 \\
21940 & 17/09/04 15:01:26 GMT & SSS17a & 5.0 & 0.147 &  &  & 3 \\
21940 & 17/09/05 15:31:29 GMT & SSS17a & 5.0 & 0.162 &  &  & 3 \\
21940 & 17/09/10 14:54:42 GMT & SSS17a & 5.0 & 0.126 &  &   & 3 \\
21940 & 17/09/11 14:44:42 GMT & SSS17a & 5.0 & 0.151 &  &  & 3 \\
21940 & 17/09/12 14:03:28 GMT & SSS17a & 5.0 & 0.113 &  &  & 3 \\
21940 & 17/09/14 13:26:42 GMT & SSS17a & 5.0 & 0.147 &  &  & 3 \\
21940 & 17/09/15 14:03:28 GMT & SSS17a & 5.0 & 0.106 &  &  & 3 \\
21940 & 17/09/16 14:03:28 GMT & SSS17a & 5.0 & 0.118 &  &  & 3 \\
21940 & 17/09/17 14:03:28 GMT & SSS17a & 5.0 & 0.111 &  &  & 3 \\
21940 & 17/09/18 14:03:28 GMT & SSS17a & 5.0 & 0.111 &  &  & 3 \\
21940 & 17/09/21 14:18:28 GMT & SSS17a & 5.0 & 0.132 &  &  & 3 \\
21940 & 17/09/22 13:34:42 GMT & SSS17a & 5.0 & 0.121 &  &  & 3 \\
21975 & 17/08/17 20:00:00 GMT &  & 0.025 & 200000.0 &  &  & NA \\
21975 & 17/08/17 20:00:00 GMT &  & 0.045 & 100000.0 &  &  & NA \\
29053 & 20/12/16 01:40:37 GMT &  & 3.0 & 0.013 &  &  & 3 \\
32094 & 22/05/23 10:46:25 GMT &  & 3.0 & 0.006 &  & & 4    
\end{longtable*}

In addition to \texttt{GPT-4.1} we experimented with other LLMs, such as \texttt{Llama3.1-405B-Instruct}. 
To give an apples-to-apples comparison between \texttt{GPT-4.1} and \texttt{Llama3.1-405B-Instruct}, we use the latter to evaluate its performance using the same used in the \texttt{GPT-4.1}-powered GCN parser. We present the performance metrics for \texttt{Llama 3.1 405B-Instruct} in Table~\ref{tab:GCNparser_metric_llama}. These performance metrics may be improved by optimizing the prompts for use with \texttt{Llama3.1-405B-Instruct}. 

\begin{table}[htbp]
    \begin{center}
    \caption{Performance metrics of the resilience of the AI-powered GCN parser, which integrates \texttt{Llama3.1-405B-Instruct} as its language modeling component. Metrics include precision, recall, and F1 score for event matching, as well as GCN-specific recall (GCN-R), reported as mean values with standard deviations across test conditions.
    \label{tab:GCNparser_metric_llama}
    }
    \begin{tabular}{lcc}
    \hline 
    \hline
    Metric & Mean & Standard Deviation \\
    \hline
    Event match precision, $P$ & 0.512 & 0.022 \\
    Event match recall, $R$ & 0.447 & 0.019 \\
    Event match F1 & 0.477 & 0.021 \\
    GCN match recall, GCN-R & 0.375 & 0.018 \\
    \hline
    \end{tabular}
    \end{center}
\end{table}

\bibliography{main}{}
\bibliographystyle{aasjournal}

\end{document}